\documentclass[useAMS,usenatbib,a4paper]{mn2e}
\usepackage{amsmath}
\usepackage[pdftex]{graphicx}
\usepackage{epstopdf}
\usepackage[english]{babel}
\usepackage{subfigure}
\usepackage{array,multirow}
\usepackage{underscore}
\usepackage{natbib}
\usepackage{float}
\usepackage{amssymb}
\usepackage{placeins}
\usepackage{tabularx}
\usepackage{caption}

\newcolumntype{L}[1]{>{\raggedright\let\newline\\\arraybackslash\hspace{0pt}}m{#1}}
\newcolumntype{C}[1]{>{\centering\let\newline\\\arraybackslash\hspace{0pt}}m{#1}}
\newcolumntype{R}[1]{>{\raggedleft\let\newline\\\arraybackslash\hspace{0pt}}m{#1}}

\def\mnras{MNRAS}
\def\apj{ApJ}

\def\apjl{ApJL}
\def\apjs{ApJS}
\def\araa{ARA\&A}

\def\nat{Nature}

\def\procspie{Proc. SPIE}

\title[M87's radially varying IMF]{Galaxy structure from multiple tracers: III. Radial variations in M87's IMF}
\author[L. J. Oldham \& M. W. Auger]{Lindsay Oldham$^1$\thanks{E-mail: lindsay.oldham@cfa.harvard.edu}\thanks{Menzel Fellow} \& Matthew Auger$^2$\\
$^{1}$ Harvard-Smithsonian Center for Astrophysics, 60 Garden Street, Cambridge, MA 02138, USA \\
$^{2}$ Institute of Astronomy, University of Cambridge, Madingley Road, Cambridge CB3 0HA, UK \\
}

\begin{document}
\maketitle
\setcounter{page}{1}

\begin{abstract}
\noindent
We present the first constraints on stellar mass-to-light ratio gradients in an early-type galaxy (ETG) using multiple dynamical tracer populations to model the dark and luminous mass structure simultaneously. We combine the kinematics of the central starlight, two globular cluster populations and satellite galaxies in a Jeans analysis to obtain new constraints on M87's mass structure, employing a flexible mass model which allows for radial gradients in the stellar mass-to-light ratio. We find that, in the context of our model, a radially declining stellar-mass-to-light ratio is strongly favoured. Modelling the stellar mass-to-light ratio as following a power law, $\Upsilon_{\star} \sim R^{-\mu}$, we infer a power-law slope $\mu = -0.54 \pm 0.05$; equally, parameterising the stellar-mass-to-light ratio via a central mismatch parameter relative to a Salpeter IMF, $\alpha$, and scale radius $R_M$, we find $\alpha > 1.48$ at $95\%$ confidence and $R_M = 0.35 \pm 0.04$ kpc. We use stellar population modelling of high-resolution 11-band HST photometry to show that such a steep gradient cannot be achieved by variations in only the metallicity, age, dust extinction and star formation history if the stellar initial mass function (IMF) remains spatially constant. On the other hand, the stellar mass-to-light ratio gradient that we find is consistent with an IMF whose inner slope changes such that it is Salpeter-like in the central $\sim 0.5$ kpc and becomes Chabrier-like within the stellar effective radius. This adds to recent evidence that the non-universality of the IMF in ETGs may be confined to their core regions, and points towards a picture in which the stars in these central regions may have formed in fundamentally different physical conditions. 
\end{abstract}


\section{Introduction}
\label{sec:chap7sec1}

The distribution of masses with which stars form is a fundamental property of a galaxy, and has an impact on virtually everything that we subsequently observe.  However, the nature of the IMF in environments beyond our Milky Way remains uncertain. Whilst the IMF appears to be strikingly uniform across the diversity of environments within our own Galaxy \citep{Bastian2010}, and adequately described by a simple broken power law \citep{Kroupa}, recent years have brought to light an accumulation of evidence that the same may not be true extragalactically. Independent techniques based on strong gravitational lensing and stellar kinematics \citep[e.g.][]{Auger2010b,Cappellari2012} have indicated that more massive ETGs have more mass in stars than is predicted by a Milky-Way-like IMF, which the analysis of stellar-surface-gravity-sensitive spectral features has attributed to an excess of low-mass stars \citep{vanDokkum2010a}; this suggests a scenario in which the IMFs of more massive galaxies are more \textit{bottom-heavy} than that of the Milky Way.

However, the astrophysical processes underlying these results remain extremely uncertain. The observed size evolution of ETGs \citep[e.g.][]{Daddi2005,vanDokkum2010b} supports the idea that these systems grow significantly over time, primarily via minor mergers and accretion \citep{Naab2009, Hopkins2009}. If the IMF is non-universal, then the link between the formation conditions of the first stellar populations and the IMF at $z=0$ is complicated by the fact that the IMFs of the stellar populations formed \textit{in situ} and those that were accreted may differ. Moreover, the interpretation of the observed variations of the IMF as a function of galaxy velocity dispersion \citep{Treu2010} is further complicated by (a) the degeneracy between dark and stellar mass, which so far has had to be broken by assuming a simple form for the halo (which was shown by \citealp{Auger2010b} to affect the strength of the correlation that is inferred), and (b) the fact that the calculation of global IMF mismatch parameters (see Equation~\ref{eq:IMF}) depends on luminosity-weighted properties integrated over some aperture (e.g. the Einstein radius of a lens, or a spectral aperture), introducing a non-uniformity between measurements and making it difficult to interpret trends across the galaxy population quantitatively. 

Recently, a key step towards overcoming these limitations was provided by \citet{MartinNavarro2015}, where gravity-sensitive spectral features were analysed as a function of radius for three nearby ETGs. The result indicated that the two high-mass ETGs exhibit significant radial IMF gradients -- with bottom-heavy IMFs in their central regions, which become Milky-Way-like at larger radii -- whilst the IMF of the lower-mass system is consistent with being Milky-Way-like at all radii. In the context of the two-phase scenario of ETG formation -- in which a compact core forms at early times, followed by lower-density wings due to the accretion of lower-mass satellites \citep[e.g.][]{Naab2009} -- this result points towards a picture in which the initial star formation processes in the progenitors of ETG cores are fundamentally different from those in lower-mass galaxies. 

However, whilst stellar population studies such as this can suggest a radial dependence of the fraction of low-mass stars -- and therefore the low-mass end of the IMF -- they cannot formally provide any information about the high-mass end of the IMF. This must be investigated using probes such as dynamics and strong gravitational lensing, which are sensitive to the \emph{total} stellar mass-to-light ratio $\Upsilon_{\star}$, or the IMF \emph{normalisation}, rather than the IMF \emph{shape}. The first dynamical study of this kind, however -- which used molecular gas kinematics to dynamically trace the stellar mass-to-light ratio in the inner 1-2 kpc of seven massive ETGs -- found a large scatter in both the overall IMF normalisation (ranging from sub-Chabrier to super-Salpeter) and the slope of the radial profile (including rising, falling and flat profiles), which furthermore did not seem to correlate with any global galaxy properties \citep{Davis2016}. 

One problem with focusing exclusively on \emph{central} kinematics (i.e. gas or stars) is that the mass contribution from the dark matter halo cannot be well constrained, and must consequently be either subject to strong assumptions or ignored, which adds significant uncertainty to the resulting measurement of any stellar mass-to-light ratio gradient. To make progress, more extensive modelling must be carried out in order to infer the stellar mass and the dark halo structure simultaneously. However, these two mass components are extremely degenerate and can only be robustly separated, in the context of a model, by combining multiple mass probes, each of which independently measures the gravitational potential (see also \citealp{Dutton2013b,Smith2015} for methods to constrain the dark matter using statistical galaxy samples and simulations, respectively). In \citet{Oldham2016b} we combined multiple independent dynamical tracers of the potential in the giant elliptical M87, the brightest cluster galaxy (BCG) in Virgo, to infer the black hole mass, the structure of the dark matter halo and the stellar mass-to-light ratio -- which was assumed to be uniform across the galaxy -- and found that M87 could be best described by a centrally cored dark halo but that the inference on the IMF was dependent on our assumptions about the stellar orbital anisotropy. Here, we overcome this limitation using updated kinematic data and a more flexible model to extend the previous analysis to investigate the possibility of radial gradients in the stellar mass-to-light ratio for the first time. 

The paper is structured as follows. In Sections~\ref{sec:chap7sec2},~\ref{sec:chap7sec3} and~\ref{sec:chap7sec4} we introduce the data, our dynamical modelling and our statistical modelling, the main results of which we present in Section~\ref{sec:chap7sec5}. Section~\ref{sec:chap7sec6}  compares our dynamical constraints with expectations from stellar population modelling; we then discuss our findings in Section~\ref{sec:chap7sec7} and summarise in Section~\ref{sec:chap7sec8}. All magnitudes are in the AB system and the distance to M87 is assumed to be $D_L = 16.5$ Mpc.

\section{Data}
\label{sec:chap7sec2}

To disentangle the contributions to M87's gravitational potential from the stellar mass and the dark matter, we require the kinematics of multiple independent tracer populations to simultaneously satisfy the Jeans equation for the same potential. Here, we use the kinematics of stars, globular clusters and satellite galaxies, which span a radius range from $\sim$10 pc to 1 Mpc. In the following sections, we summarise the data used to characterise these different tracer populations.

\subsection{Photometry}
\label{sec:chap7sec2sub1}

\begin{figure*}
  \centering
\includegraphics[trim=20 0 10 0,clip,width=0.49\textwidth]{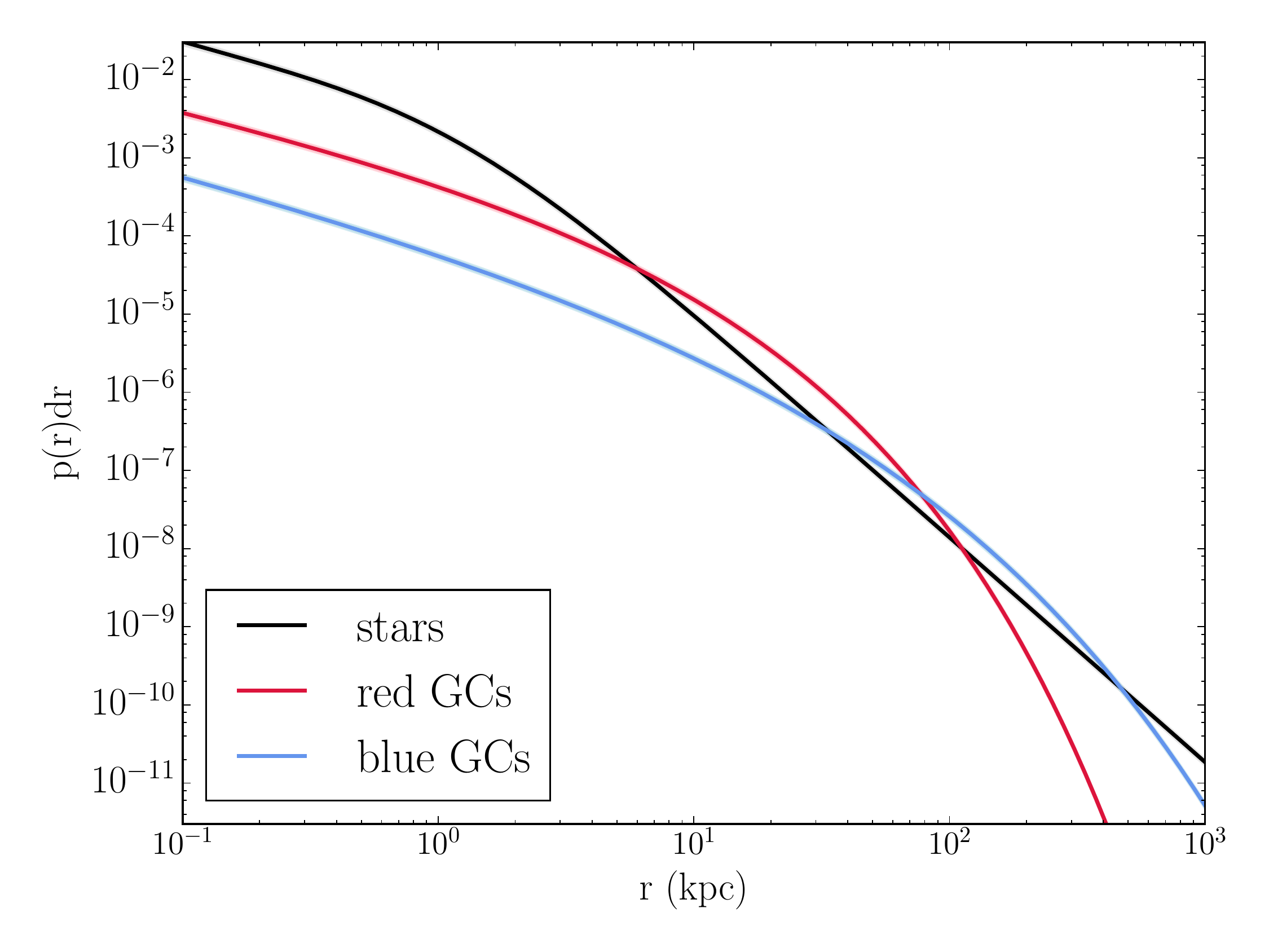}\hfill
\includegraphics[trim=20 0 10 0,clip,width=0.49\textwidth]{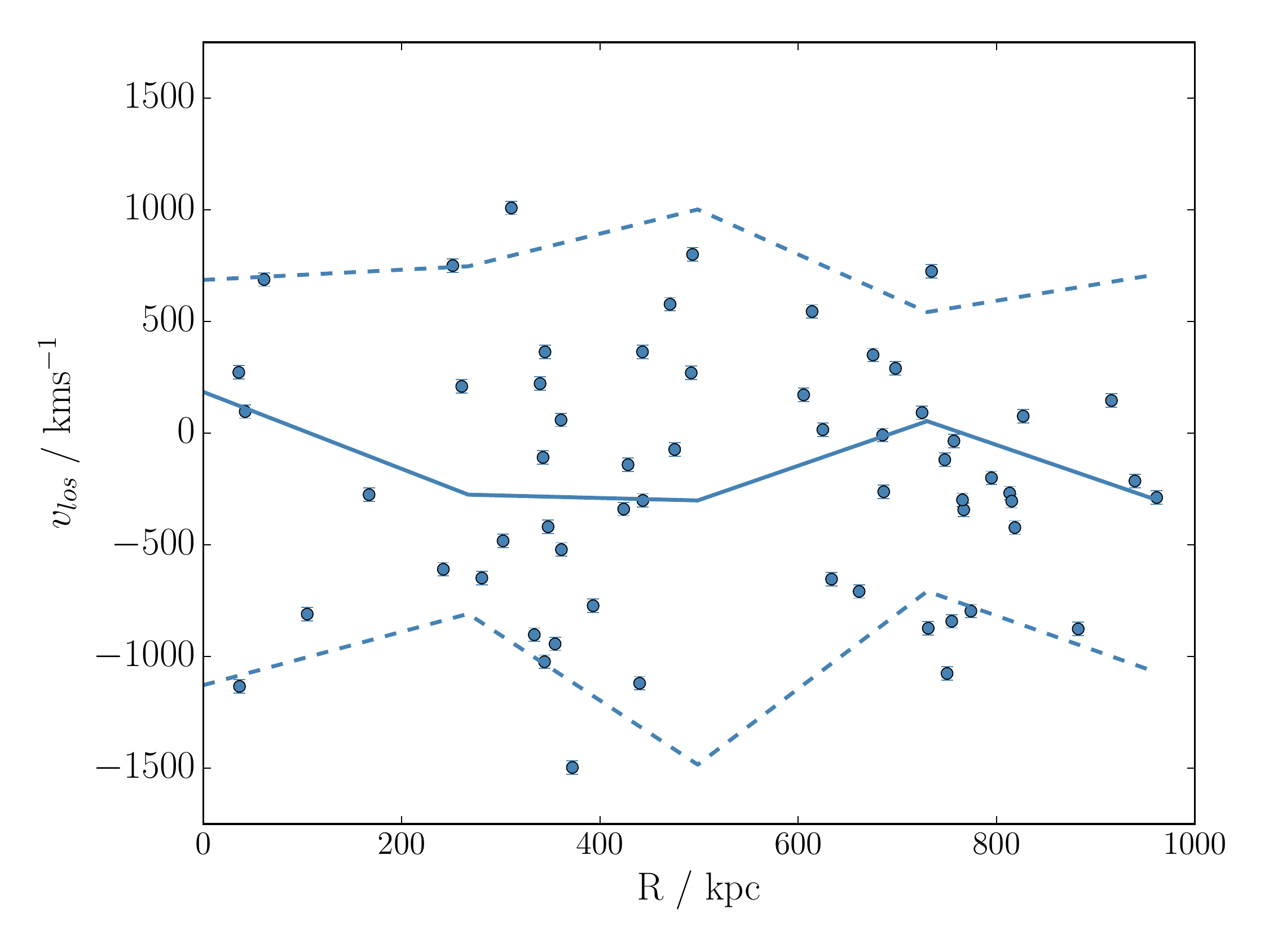}\hfill
 \caption{Left: Normalised probability distribution functions for the deprojected 3D luminosity profiles of the dynamical tracer populations as a function of radius. Uncertainties are included, but are small. The stellar profile is modelled with a Nuker profile as in Equation 1; each globular cluster population follows a S\'ersic profile in radius. Right: Line-of-sight velocity distribution of satellite galaxies as a function of projected radius, with the running median (solid line) and $3\sigma$ tracks (dashed lines) shown.}
 \label{fig:chap7fig1}
\end{figure*}

Use of the Jeans equation requires knowledge of the luminosity density $l(r)$ of each population from which kinematic tracers are drawn. This in turn depends on high-quality photometry. The datasets we use for this purpose are identical to those presented in \citet{Oldham2016b}, which should be referred to for further details. A summary is provided below and in Figure~\ref{fig:chap7fig1} (left).

For the \textbf{stellar surface brightness}, we model the radial $V$-band profile of \citet{Kormendy2009} with a Nuker profile (exactly as in \citealp{Oldham2016b}). Assuming  spherical symmetry, we deproject this profile to give the 3D luminosity density shown (with uncertainties based on our fit to the \citealp{Kormendy2009} profile) in Figure~\ref{fig:chap7fig1} (left).

For the \textbf{globular clusters}, we use the colour and radial profiles presented in \citet{Oldham2016a}, in which CFHT/MegaPrime imaging in the \textit{ugriz} bands was used to compile a globular cluster candidate catalogue. We model the distributions of the two (red and blue) globular cluster populations, in addition to those of interloping objects. The globular clusters were treated probabilistically as being drawn from the red and blue populations, each of which is described by a S\'ersic radial profile, a Gaussian luminosity function and radially-dependent Gaussian colour profiles. The 3D deprojected radial profiles for the red and blue populations are also shown in Figure~\ref{fig:chap7fig1} (left), with uncertainties as calculated in \citet{Oldham2016a}.

Finally, for the \textbf{satellite galaxies}, we do not compile a photometric sample, as we would expect this to be incomplete and the selection function intractable; we therefore incorporate this population into our model using the robust mass estimator presented in \citet{Watkins2010}, rather than a Jeans analysis. This is described in more detail in \citet{Oldham2016b} and Section~\ref{sec:chap7sec3}.

\subsection{Kinematics}
\label{sec:chap7sec2sub2}

As our dynamical model is based on solutions to the spherical Jeans equation, we combine the photometric information of Section~\ref{sec:chap7sec2sub1} with the kinematic data summarised below. The data used in this study are similar to those described in \citet{Oldham2016b}, but differ in two key respects. First, the globular cluster sample is almost doubled due to recent MMT/Hectospec observations \citep[e.g.][]{Ko2017} and samples M87's surroundings more representatively and extends to larger radii. Second, the stellar kinematics from SAURON are superseded by new, higher-signal-to-noise kinematics from MUSE. 

We combine \textbf{stellar kinematics} from two datasets which are complementary in spatial resolution and extent. In the central 2$''$ ($\sim 170$ pc), we use the velocity dispersions presented in \citet{Gebhardt2011}, which were obtained using adaptive optics on Gemini/NIFS and binned radially in bins of $\log r = 0.16$, with a spatial resolution of $\sim0.1''$ ($=8$ pc) and a signal-to-noise generally $> 50$. At larger radii, we introduce a new measurement of the 2D velocity dispersion profile obtained using VLT/MUSE \citep{Bacon2010}. The central arcminute of M87 was observed for one hour on the night of June 28 2014 during the MUSE science verification phase, and the data are available in the ESO archive. We reduced the datacube using the standard ESOREX pipeline and modelled the resulting spectra, binned to 0.6 arcsecond pixels, in the rest wavelength range 5000-5775$\textrm{\AA}$. We follow the methods of \citet{Oldham2017HD} by modelling the spectra as the linear combination of a set of stellar templates from the INDO-US library and an additive order-10 polynomial to account for the continuum. The velocity dispersion map that we obtain is consistent with that presented in \citet{Emsellem2014} and with the SAURON data at the $2\sigma$ level, though we note that our MUSE velocity dispersions rise more steeply in the central arcsecond, more closely following the NIFS kinematics in that region (this may in part be due to differences in the PSF). For the dynamical modelling, we impose a minimum uncertainty of 5\% to account for systematic uncertainties due to template choice. 

For the \textbf{globular clusters}, we update the kinematic catalogue of \citet{Strader2011} to include the new globular cluster candidates that have been observed with MMT/Hectospec and are available on the CfA archive\footnote{http://oirsa.cfa.harvard.edu/}. The original \citet{Strader2011} catalogue combines measurements for 451 globular clusters -- obtained using Keck/DEIMOS, Keck/LRIS and MMT/Hectospec -- with literature data to provide line-of-sight velocity measurements for a total of 612 globular clusters within 240 kpc of M87. To supplement this, we cross-correlate the 2391 objects within $\sim 1$ degree ($\sim 300$ kpc) of M87 with measured velocities from MMT/Hectospec with the photometric globular cluster catalogue of \citet{Oldham2016a} and select as globular cluster candidates those objects which satisfy the colour cuts that were used in that photometric study. We also impose a cut in line-of-sight velocity relative to M87 of 800 kms$^{-1}$. Finally, we require candidates to have probabilities of being a globular cluster $P(GC)$ (as opposed to an interloper; see \citealp{Oldham2016a}) satisfying $P(GC)>0.6$, giving a final sample of 900 objects. Of these, 95\% have $P(GC)>0.65$ and 65\% have $P(GC)>0.95$, matching the original \citet{Strader2011} catalogue. we verify that this selection does not bias our inference relative to the original catalogue by rerunning our isotropic inference without the new globular cluster velocities, and find that the results are consistent within $2\sigma$. The median uncertainty on the globular cluster velocities is $\sim 15\%$ (due to a long tail at the large-uncertainty end of the distribution), which is approximately double that of the MUSE kinematics.


For the \textbf{satellite galaxies}, we use the same catalogue of 60 Virgo galaxies as in \citet{Oldham2016b}. Full details of the selection are explained in that paper, but in summary, we cross-correlated the Extended Virgo Cluster Catalogue (EVCC), which includes line-of-sight velocities and sky positions for 1589 candidate cluster members \citep{Kim2014}, with the distance modulus catalogue of \citep{Blakeslee2009} to select galaxies (a) classified as `certain' cluster members by both of the criteria used in the EVCC and (b) with luminosity distances $D_L < 20$ Mpc. We further imposed a declination angle cut $\delta > 9.5$ deg and a cut in projected distance relative to M87 $R < 1$ Mpc. The velocity-radius distribution of the resulting catalogue is shown in Figure~\ref{fig:chap7fig1} (right).

\section{Dynamical model}
\label{sec:chap7sec3}

We construct models for M87's mass density and the anisotropy of each tracer population using the spherical Jeans equation.

\subsection{Mass components}

Our model for the total mass density of the galaxy $\rho(r)$ consists of a dark matter halo, a stellar mass component and a black hole:
\begin{equation}
 \rho(r) = \rho_{DM}(r) + \rho_{\star}(r) + \rho_{BH}(r).
\end{equation}
Guided by the results of \citet{Oldham2016b}, in which a number of different models for the dark halo were investigated, we use a generalised Navarro-Frenk-White (gNFW) profile to describe the dark matter halo
\begin{equation}
 \rho_{DM}(r) = \frac{\rho_0}{4\pi}\Big(\frac{r}{r_s}\Big)^{-\gamma} \Big(1 + \frac{r}{r_s}\Big)^{\gamma-3}
\end{equation}
where the scale radius $r_s$, inner slope $\gamma$ and density scale $\rho_0$ are parameters to be inferred. In \citet{Oldham2016b}, we found that this profile was flexible enough to provide a good description of the data at relatively low computational cost.

The black hole is simply a point mass $M_{BH}$ at the galaxy centre such that
\begin{equation}
 \rho_{BH}(r) = \frac{M_{BH}}{4\pi r^2} \delta(r).
\end{equation}

\subsection{Stellar mass-to-light ratio}

The key step forward in this study with respect to the models of \citet{Oldham2016b} is that we now model the stellar mass density with a mass-to-light ratio $\Upsilon_{\star}$ that follows a radially varying profile. Thus, the projected surface mass density $\Sigma_{\star}(R)$ is related to the surface brightness distribution $I_{\star}(R)$ as
\begin{equation}
 \Sigma_{\star}(R) = \Upsilon_{\star}(R) I_{\star}(R)
\end{equation}
for projected radius $R$. To assess the sensitivity of our inference to the assumed form of $\Upsilon_{\star}(R)$, we consider two profiles. Firstly, we consider a \textbf{power-law (PL) profile} in which
\begin{equation}
\log \Upsilon_{\star}(R) = \log \Upsilon_{\star,1} + \mu \log R,
\end{equation}
with power-law index $\mu$, and $\Upsilon_{\star,1}$ representing the stellar mass-to-light ratio at a projected distance of 1 kpc from the galaxy centre. In this model, however, the stellar mass becomes unphysically large in the very centre, which prevents us from making meaningful inference on the black hole mass; we therefore fix $M_{BH} = 6.6 \times 10^{9} M_{\odot}$ to be consistent with the inferences of \citet{Gebhardt2011} and \citet{Oldham2016b}.

In reality, however, there exists some covariance between $M_{BH}$ and $\Upsilon_{\star}(R)$ that is important to explore; we therefore consider a second \textbf{Salpeter-to-Chabrier (SC) model} in which $\Upsilon_{\star}$ tends to a finite value centrally and becomes Chabrier-like at large radii:
\begin{equation}
 \Upsilon_{\star}(R) = \alpha_s \Upsilon_{\star,s} + \frac{\Upsilon_{\star,ch} - \alpha_s \Upsilon_{\star,s}}{R^2 + R_M^2}R^2
\end{equation}
for stellar mass-to-light ratios assuming Salpeter and Chabrier IMFs $\Upsilon_{\star,s}$, $\Upsilon_{\star,ch}$ inferred from photometry (see Section~\ref{sec:chap7sec6}), mismatch parameter relative to a Salpeter IMF $\alpha_s$ and 2D scale radius $R_M$. Thus $\Upsilon_{\star}(R \to 0) = \alpha_s \Upsilon_{\star,s}$ and $\Upsilon_{\star}(R \to \infty) = \Upsilon_{\star,ch}$. As $\alpha_s$ can be either greater or less than unity, this model makes no assumptions about whether $\Upsilon_{\star}$ rises or falls with radius, and the free scale radius $R_M$ allows for the possibility that $\Upsilon_{\star}$ is not Chabrier-like at any radii (i.e. $R_M$ is allowed to become large). In this model, we allow the black hole mass to be a free parameter. We also explore generalisations of the SC model in which the index to which the projected radius is raised is a free parameter and the outer asymptotic stellar mass-to-light ratio is allowed to vary (i.e. $\Upsilon_{\star,ch}$ becomes $\alpha_{ch}\Upsilon_{\star,ch}$); however, we find that our constraints on these extra parameters are weak and that their inclusion does not significantly affect our results.


\subsection{Anisotropy}

To assess the robustness of our inference against different assumptions regarding the orbital anisotropy of the tracers, we also consider two different anisotropy models. Firstly, we consider an \textbf{isotropic} model, in which all tracers have zero anisotropy, $\beta = 0$, at all radii. This is the simplest assumption that can be made here, and is motivated by previous dynamical studies of M87 \citep[e.g.][]{Zhu2014,Cote2001,Murphy2011}, which have found the orbital structure of the globular cluster population and the stars to be nearly isotropic. However, we emphasise that the anisotropy structure of stars, globular clusters and satellite galaxies in ETGs is not well understood from either observations- or simulations-based viewpoints, and that our assumptions about the anisotropy may be an important source of additional uncertainty in our analysis; this is discussed further in Section 5.4. In our main analysis, we mitigate this uncertainty by additionally considering a more sophisticated \textbf{anisotropic} model in which the stars follow a scaled Osipkov-Merrit profile
\begin{equation}
 \beta_{\star} = \beta_{\star,0} \frac{r^2}{r^2 + r_a^2}
\end{equation}
where the scale radius $r_a$ and asymptotic anisotropy $\beta_{\star,0}$ are parameters to be inferred. Since the globular cluster kinematics are sparser, we cannot constrain such a complex model for their anisotropies, and instead treat them as having constant anisotropies $\beta_{r,b} =$~constant, where the subscripts $r$, $b$ refer to the red and blue populations respectively. Since these two populations are dynamically independent, their anisotropies are also treated as such.

\subsection{Large-radius mass}

As noted in Section~\ref{sec:chap7sec2sub1}, the underlying luminosity distribution of the satellite galaxies is not well understood; the catalogues from which our data are drawn were selected spectroscopically and are thus almost certainly subject to some unknown selection function. We therefore incorporate this population into our dynamical model using the mass estimator presented by \citet{Watkins2010}, which we calibrate using massive haloes from the MultiDark simulation \citep{Prada2012}. This procedure is detailed in \citet{Oldham2016b}, but in summary, we select all haloes with more than 30 subhaloes in the $z=0$ halo catalogue of the first MultiDark simulation, and use the subhalo positions and velocities and the parent halo mass profiles to infer the posterior distributions on the subhalo density slope $\nu$ and parent halo potential slope $\mu$ -- both assumed to be scale-free -- and the subhalo anisotropy $\beta$ for the population, which are required for use of the mass estimator. We use this inference to generate a posterior on the total mass of M87 within the outermost satellite in our catalogue, whose distribution we use to reselect haloes from the simulation, and iterate the procedure until our inferences on $\mu$, $\nu$ and $\beta$ converge. 

We test the calibration by applying the resulting mass estimator to the MultiDark parent haloes, and find that we recover their masses with a negligible median offset and a scatter of 0.1 dex (as shown in Figure 3 of \citealp{Oldham2016b}). When applied to M87, this gives a constraint on the total mass within the projected radius of the outermost satellite galaxy:
\begin{equation}
 \log\Big(\frac{M(R<985\text{kpc})}{M_{\odot}}\Big) = 14.11 \pm 0.19
\end{equation}
which allows us to normalise the mass profile appropriately at large radii. In \citet{Oldham2016b}, we showed that the removal of this constraint from our inference did not change our results within the uncertainties; nevertheless, the same is not necessarily true for this study, in which new data and a more complex mass model are used, and we therefore retain it in the analysis (but again investigate its importance in Section 5.2).

\section{Statistical model}
\label{sec:chap7sec4}

We compare the stellar line-of-sight velocity dispersions calculated from the Jeans equation directly with the measured line-of-sight velocity dispersions, giving a contribution to the likelihood
\begin{equation}
 \ln L_{\star,k} = -\frac{1}{2}\Big(\frac{\sigma_k-\sigma_m}{\delta_k}\Big)^2 - \frac{1}{2} \ln \big(2 \pi \delta_{\sigma_k}^2\big)
\end{equation}
for the $k^{th}$ stellar velocity dispersion measurement. Here, $\sigma_k$ and $\sigma_m$ represent the observed and model velocity dispersions respectively, and the uncertainty $\delta_k$ is the quadratic sum of the measured uncertainty $\Delta_{k}$ and a regularisation term $\Delta_{\star}$. For the NIFS dataset, which is small, we set $\Delta_{\star} = 0$, whereas for the MUSE dataset, we infer $\Delta_{\star}$ to allow the stellar and globular cluster datasets to select their preferred relative weighting and to prevent the MUSE data from over-dominating the inference. We verify that this is a reasonable step by checking that our inferred value of $\Delta_{\star}$ is consistent with the scatter between the data and our best model in the left-hand panel of Figure 3.

We compare the model globular cluster velocity dispersions with the observed globular cluster velocities by modelling the velocity distribution of each globular cluster population as a Gaussian with the dispersion calculated from the Jeans equation. We then assign probabilities of belonging to either the red or the blue population to each globular cluster candidate, based on its velocity, colour, magnitude and spatial information, and stochastically sample these probabilities at each step in our MCMC exploration. At any step in this stochastic sampling, the likelihood contribution from the $k^{th}$ globular cluster is
\begin{equation}
 \ln L_{GC,k} = -\frac{1}{2}\frac{v_k^2}{\delta v_k^2 + \sigma_m^2} - \frac{1}{2} \ln \big(2 \pi (\delta v_k^2 + \sigma_m^2)\big)
\end{equation}
for globular cluster velocity $v_k$, measurement uncertainty $\delta v_k$ and model velocity dispersion $\sigma_m$, which is that of either the red or the blue population depending on the globular cluster's classification at that sampling step.

For the satellite galaxies, we compute the mass enclosed within the radius of the outermost object and compare it with the mass calculated from the mass estimator, giving a single contribution to the likelihood
\begin{equation}
 \ln L_{sat} = -\frac{1}{2}\Big(\frac{\log M_{sat} - \log M_{m}}{\delta \log M_{sat}}\Big)^2 - \frac{1}{2} \ln (2 \pi \delta \log M_{sat}^2)
\end{equation}
for model mass $\log M_{m}$, and the mass and uncertainty calculated from the mass estimator, $\log M_{sat} = 14.11$, $\delta \log M_{sat} = 0.19$, measured within $R_{out} = 985$ kpc.

For each of the isotropic and anisotropic runs, we explore the parameter space using \textsc{Emcee} \citep{ForemanMackey2013}.

\section{Assessment of model assumptions}

Our dynamical model necessarily makes a number of assumptions, namely: (a) spherical symmetry, (b) dynamical equilibrium, and (c) anisotropy structure. In this section, we consider and quantify the significance of each.

\subsection{Spherical symmetry}

The stellar and globular cluster distributions in M87 appear round in the central regions (see Figure 1 of \citealp{Zhu2014}), with $q > 0.9$ for $R < 10$ kpc, but become flatter at larger radii (\citealp{Zhu2014} measure axis ratios $q \sim 0.6$ for the globular clusters and $q \sim 0.7$ for the stars at $R \sim 100$ kpc). Equally, simulations have shown that the shape of the dark halo may be correlated with -- but generally rounder than -- that of the luminous matter \citep{Wu2014}. It is therefore possible that our use of the \emph{spherical} Jeans equation may lead to a bias in our inference on the mass. However, \citet{Zhu2014} investigated this possibility by comparing axisymmetric Jeans models for M87 involving different assumptions about the flattening of the globular cluster distribution and halo mass, and found negligible scatter in the inference on the mass profile within $R \lesssim 60$ kpc, and a maximum mass difference at large radii between different models of $\sim$10\%. Since the models agree closely at small radii, where the stars dominate, and only begin to deviate on scales where the dark mass far exceeds the luminous mass, we do not consider this to add significant uncertainty to our determination of $\Upsilon_{\star}(R)$, but rather impose a 10\% systematic uncertainty on the normalisation of the dark matter halo, $\rho_0$.

\subsection{Dynamical equilibrium}

In carrying out a Jeans analysis, we are assuming that the stars and globular clusters are in dynamical equilibrium; since the mass estimator of \citet{Watkins2010} is derived from the Jeans equations, this assumption holds for the satellite galaxies as well. However, it is possible that a number of the globular clusters and satellite galaxies at large radii have been accreted during minor mergers and have not yet relaxed into an equilibrium configuration or may be associated with tidal streams. Indeed, \citet{Romanowsky2012} found evidence for an inner shell of globular clusters around $R \sim 50 - 100$ kpc and an outer stream of globular clusters around $R \sim 170$ kpc, both potentially left behind by past merger events. However, the fraction of globular clusters associated with such structures is small: \citet{Romanowsky2012} report $15 \pm 10$ globular clusters associated with the outer stream, and emphasise that the broad spread of velocities of the majority of globular clusters implies that they are dynamically relaxed. Indeed, the phase space distribution of our globular cluster sample, shown in the right-hand panel of Figure 3, does not exhibit any abrupt features, further confirming that it is dominated by dynamically relaxed objects. The impact of these structures on our inference should therefore also be small. 

The larger radial range of the satellite galaxies might lead us to expect a higher fraction of the satellite galaxy sample to be still falling in towards the cluster centre and thus potentially out of equilibrium. Indeed, \citet{OldhamEvans} found tentative evidence for clustering in the satellite galaxy population (though that result is highly model-dependent). Any bias due to our equilibrium assumption for this population could therefore be significant. We investigate this bias by rerunning our isotropic inferences with the uncertainty on the satellite galaxy constraint doubled, and, similarly to \citet{Oldham2016b}, find that our inference on the density structure does not change within the uncertainties. Essentially, the contribution of this population to the likelihood is much smaller than that of the other tracer populations -- for which more extensive data are available -- such that the satellite mass estimate serves only to reject models which extrapolate to extremely high or low masses at large radii, but otherwise has little constraining power. Any bias introduced by incorrect assumptions about the dynamical state of these objects therefore is not a significant source of uncertainty in our analysis.

\subsection{Parametric models}

Our separation of the dark and luminous mass components depends on the parametric profiles assumed for each. In \citet{Oldham2016b}, we assumed the stellar mass to follow the light (i.e. $\Upsilon_{\star} = $const.) and, motivated by simulations, investigated a number of different parameterisations of the dark matter halo. Indeed, the combination of $\Upsilon_{\star} = $const. models with NFW or gNFW halo profiles is common practice for the mass modelling of ETGs, and, in this context, we found the requirement of the data for a core in the halo to be robust against reasonable changes to our parameterisation of the halo profile. 

In this study, we further generalise this paradigm to identify departures of $\Upsilon_{\star}$ from spatial uniformity, and this potentially introduces further degeneracies between the inner halo slope and the gradient of $\Upsilon_{\star}$. Furthermore, unlike for the halo, we are not able to motivate profiles for $\Upsilon_{\star}(R)$ based on existing simulations or previous observational constraints. It therefore becomes important to understand how strongly our conclusions depend on our choice of parameterisation for $\Upsilon_{\star}(R)$.

Our use of two different $\Upsilon_{\star}(R)$ profiles, as presented in Section 3.2, is motivated by this need. However, as discussed further in Section 6, we also consider some yet more flexible models to test the robustness of the qualitative aspects of our inferences. In our most general treatment, we remodel the stellar surface brightness profile using a multi-Gaussian expansion (MGE; \citealp{Cappellari2002}), and model the stellar mass surface density $\Sigma_{\star}$ using the same expansion, but allowing the amplitudes of the individual Gaussian components to vary independently. The ratio of the inferred $\Sigma_{\star}(R)$ to the fitted $I_{\star}(R)$ then gives a stellar-mass-to-light ratio which is relatively free from assumptions about its radial structure. For instance, with this model, it is possible for $\Upsilon_{\star}(R)$ to vary non-monotonically, and the large- and small-radius behaviours are not correlated as they are in our default models (note that this means that the $\Upsilon_{\star}(R)$ profile is also more weakly constrained). The result of this experiment is presented in Section 6, but essentially, we find that this model reproduces the general behaviour of our default models. Indeed, due to our prior ignorance about the form of $\Upsilon_{\star}(R)$, the primary aim of this study is to understand the stellar-mass-to-light ratio structure at a qualitative level (i.e. whether it is consistent with being flat, or rising or declining as a function of radius), and the result of this experiment suggests that this aim -- at least in the context of our halo model -- is realistic.

On the other hand, it is possible that shortcomings in our halo model may also contribute to our inference on the stellar mass structure. In \citet{Oldham2016b}, we considered four different halo profiles and found that our inference on the spatially constant stellar-mass-to-light ratio was not dependent on our choice of halo model; though this motivated our adoption of a single (gNFW) halo profile in this study, it is not necessarily the case that our new, more flexible model for the stellar mass structure is equally insensitive to our choice of halo profile. To probe the degeneracy between the halo inner slope $\gamma$ and the stellar-mass-to-light ratio gradient $\mu$, we carry out additional inferences in which the halo slope $\gamma$ is fixed to different values, and again find that the qualitative nature of our results is unchanged. (Figures A1-A4 also show that the degeneracies between these parameters are, in practice, small.) This is likely because the mass contribution of the dark matter in the central regions is small. That a yet more flexible halo profile (e.g. a triple power law) would give rise to greater degeneracy -- and potentially remove the need for strong stellar-mass-to-light ratio gradients -- remains a possibility, which we do not explore here. Developing more flexible models for the halo structure will be an important aspect of investigations of stellar-mass-to-light ratio gradients in ETGs in the future.

\subsection{Anisotropy structure}

Finally, as discussed in Section 3.3, the well-known mass-anisotropy degeneracy, coupled with the fact that only very broad theoretical expectations for the orbital structure of ETGs exist, makes our assumptions about the anisotropy profile the largest source of uncertainty in our analysis. We have attempted to probe this degeneracy, as far as allowed by the data, by considering both isotropic and anisotropic models, and by combining multiple independent tracer populations with different characteristic radii, each of which probes the same gravitational potential but has different density and orbital structures, and a different effective radius within which it provides a robust and unbiased constraint on the total enclosed mass \citep[the so-called pinch radius of the population; see][]{Walker2009,Wolf2010,Campbell2017}. Equally, the fact that our inferences on the anisotropy (finding all tracers to have mildly radial, near-isotropic orbits; see Section 6) are consistent with previous studies of M87 using orbit-based modelling in which much weaker assumptions about the orbital structure were required \citep[e.g.][]{Murphy2011,Zhu2014}, seems to imply that our parameterisations of the anisotropies are sufficiently flexible to allow us to describe the dynamics of this system. Nevertheless, our parameterisations are clearly simplifications of the true orbital structure of any galaxy, and it is important to investigate and quantify the bias that this must introduce. 

The recent study of \citet{Read2017} used mock \textsc{Gaia} Challenge data to test a spherical Jeans modelling paradigm similar to our own, specifically considering the case of two independent (stellar) tracer populations. They found that, in the context of two stellar populations with comparable effective radii ($R_e \sim 0.5 - 1$ kpc), each with $\sim$ 60 bins and $\sim$ 60 stars per bin, they are able to precisely recover the total density profile in the region between the effective radii $R_{e,1}$, $R_{e,2}$ of the two populations, and also obtain tight constraints in the region $(0.5R_{e,1} < R < 2R_{e,2})$. However, the quality of their constraints on the central density slope depends on the true density structure: they find that they can recover central cores, but over-predict the steepness of haloes that are centrally cusped. On the other hand, their constraints on the anisotropies of the two populations are poor: though they fall within the 95th percentile confidence intervals within the effective radii, they deviate dramatically from truth at larger radii. This seems to suggest that, whilst our inference on the total mass should be unbiased, our inference on $\beta(r)$ is likely to be unreliable. We therefore treat $\beta(r)$ effectively as a nuisance parameter, and do not attempt to interpret it physically.

On the other hand, \citet{Li2016} used galaxies from the Illustris simulation to quantify the bias and scatter that are introduced when inferring the mass structure of massive ($M_{\star} > 10^{10} M_{\odot}$) oblate galaxies using the first-order axisymmetric Jeans equation (as implemented in the Jeans Anisotropic Multi-Gaussian Expansion/JAM software of \citealp{JAM}), specifically separating out the dark and luminous mass components. They found the biases in both the total mass and the separate mass components to be small, with a bias of 1\% in the total mass within 2.5$R_e$ (the radial extent of their simulated data), and $5\%$ and $-3\%$ for the stellar and dark matter masses respectively. Due to the nature of the degeneracies between $\Upsilon_{\star}$, $\gamma$ and $f_{DM}$ (see e.g. their Figure 4), this means that a the bias on the inferred inner halo slope must also be small. On the other hand, they find the scatter between inference and truth to be larger: $~\sim 10\%$ for the total mass, and 32-51$\%$ for the stellar and dark matter masses -- though, importantly, they demonstrate that the limited resolution of their MGE decomposition of the light profile contributes significantly to their difficulty in disentangling the two components; since the \citet{Kormendy2009} photometry is well-described by our Nuker profile down to small radii, this should be a much less significant problem in our analysis. 

Similarly, the orbital anisotropy along the line of sight, $\beta_z$, is recovered with negligible bias ($\Delta \beta_z = -0.02$) but significant scatter ($\delta \Delta \beta_z = 0.11$). The main difference of our analysis is that we include multiple independent tracer populations extending to $\sim 10$ times larger radii, which should help to reduce the scatter in our results. We therefore consider the uncertainties given in \citet{Li2016} as upper bounds on the accuracy that we can achieve. We note the other main difference -- that we use spherical Jeans modelling whereas \citealp{Li2016} use axisymmetric coordinates -- is not as significant, since each study uses the Jeans formalism most suited to the galaxy in question. That is, JAM is not appropriate for M87 since it assumes that the velocity ellipsoids of the galaxy are cylindrically aligned, which is a reasonable assumption for the fast rotators considered in \citealp{Li2016}, but not for M87, which is a slow rotator. Indeed, \citealp{Li2016} show in that paper that the application of JAM to prolate, M87-like galaxies performs more poorly than for oblate ones.

These studies provide some guidance for the systematic uncertainty that may be present in our inference on the mass structure due to the mass-anisotropy degeneracy, though neither supplies a direct comparison. Guided by the similarities and differences between these studies and our own, we choose to impose 10\% uncertainties on all inferred quantities. This level of uncertainty is significantly larger than the statistical uncertainties from the inference, which are small.

\section{Results}
\label{sec:chap7sec5}

\begin{figure}
 \centering
\includegraphics[trim=20 20 20 20,clip,width=0.49\textwidth]{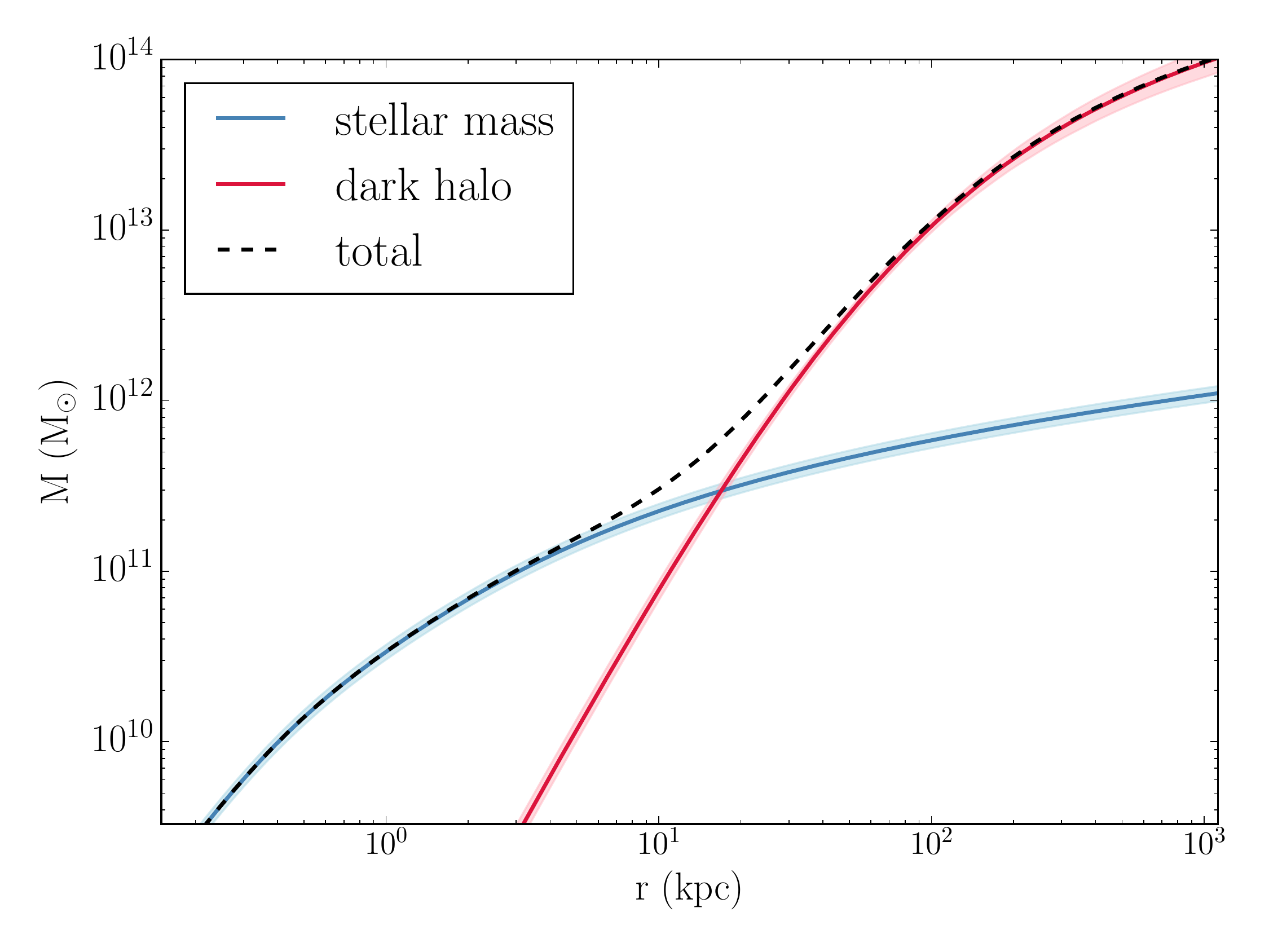}
\caption[Inferred mass profile for M87]{Inference on the dark, stellar and total mass profile in the anisotropic model. At radii $\leq 10$ kpc $\sim R_e$, the stellar mass dominates, whereas beyond this, the dark halo becomes the main contributor to the potential. Our kinematic data extend from $\sim 10$ pc to 1 Mpc, which is the radius range spanned in this Figure. Note that we include systematic uncertainties as described in Section 5.}
 \label{fig:chap7fig2}
\end{figure}

Our inferences on the structure of the dark halo and the stellar mass are shown in Figures A1 (isotropic, PL), A2 (anisotropic, PL), A3 (isotropic, SC) and A4 (anisotropic, SC) in the Appendix, and reported in Table 1. Note that, based on the discussion in the previous section, we impose minimum uncertainties of 10\% on all parameters. Our inference on the mass structure is similar in all four cases, and the halo structure we infer is also qualitatively consistent with the result of \citet{Oldham2016b}, in which a constant $\Upsilon_{\star}$ was assumed (though the size of the central dark matter core is less certain). Since the resulting stellar mass and halo profiles are similar for all models, we select the anisotropic SC model as our fiducial final model due to its increased flexibility relative to the others; the resulting mass profile, with uncertainties, is then shown in Figure~\ref{fig:chap7fig2}. For the central halo structure, we find a weak inner slope $\gamma < 0.13$ with $95 \%$ confidence and a scale radius $r_s = 60.6_{-8.3}^{+10.1}$ kpc; at large radii, the virial mass is  $\log (M_{vir}/M_{\odot}) = 14.04_{-0.12}^{+0.10}$ and the virial radius $r_{vir} = 1241_{-111}^{+103}$ kpc, consistent with our mass estimate from the satellite galaxy sample. We are also able to well reproduce the kinematics of both the stars and the globular clusters, as shown in Figure 3. Finally, we note that our inference on $\Delta_{\star} = 8.1 \pm 0.2$ kms$^{-1}$ is consistent with the scatter between our best model and the data ($\Delta \sim 9.3$ kms$^{-1}$) as shown in the left-hand panel of Figure 3; this confirms that our model is a reasonable description of the combined datasets.

The novel result from this modelling is that we can rule out a stellar mass-to-light ratio that is constant with radius with $>$99\% confidence. We find an IMF mismatch parameter $\alpha>1.48$ with 95$\%$ confidence and a small scale radius $R_M = 0.35 \pm 0.04$ kpc (quantities defined in Equation 6),  such that $\Upsilon_{\star}$ is a declining function of radius and becomes Chabrier-like by a radius of roughly 3 kpc. The stellar mass-to-light ratio profile that we infer is shown in Figure~\ref{fig:chap7fig3} (left panel), with expectations  from stellar population synthesis modelling overlaid (see Section~\ref{sec:chap7sec6}). Our PL models agree with this result: for the anisotropic PL model, we find a power-law index $\mu = -0.55 \pm 0.05$ and $\Upsilon_{\star,1} = 4.45 \pm 0.45$ (quantities defined in Equation 5), indicative of a stellar-mass-to-light ratio that is high centrally but declines with increasing radius. We confirm that the large-radius stellar-mass-to-light ratio is consistent with a Chabrier IMF by running an SC model in which the outer asymptotic stellar mass-to-light ratio is also allowed to vary as $\Upsilon_{\star} = \alpha_{ch}\Upsilon_{\star,ch}$ and find that $\alpha_{ch}$ is consistent with unity. We additionally verify the robustness of the result that the stellar-mass-to-light ratio is a declining function of radius by running a model in which the stellar light profile is described by an MGE with the amplitudes of the individual components of the MGE allowed to vary, as detailed in Section 5.3, and find that we recover a sharp decline in $\Upsilon_{\star}$ within the central 2 kpc.

Attributing this gradient wholly to a changing IMF indicates a relatively steep decline from a bottom-heavy Salpeter-like IMF in the central $\sim 0.5$ kpc to a Milky-Way-like Chabrier IMF at slightly larger radii (but still well within the stellar effective radius, and within the coverage of the MUSE kinematics). Figure~\ref{fig:chap7fig3} (right panel) emphasises this decline by recasting the stellar mass-to-light ratio in terms of the IMF mismatch parameter
\begin{equation}
\alpha_{chab} = \frac{\Upsilon_{\star,dyn}(R)}{\Upsilon_{\star,chab}(R)}
\label{eq:IMF}
\end{equation}
where $\Upsilon_{\star,dyn}(R)$ is the stellar mass-to-light ratio inferred from dynamics and $\Upsilon_{\star,chab}(R)$ is that inferred from stellar population modelling assuming a Chabrier IMF. (Note that this shows the mismatch parameter $\alpha_{chab}$ at a \emph{particular projected radius} $R$, as opposed to the mismatch that would be measured \emph{within an aperture} of radius $R$.) We note that all our models predict a sharp decline of $\Upsilon_{\star}$ within R $\sim 1$ kpc. This is surprising as it implies that the stellar population properties -- and so, potentially, the star formation conditions -- may differ significantly between the central regions and the rest of the galaxy. Of course, the stellar-mass-to-light ratio gradient that we infer may be due to gradients in the stellar age and metallicity in addition to those in the IMF; this is a possibility that we investigate in more detail in the following section.

We note, at this stage, that the statistical uncertainties from our Bayesian analysis are very small. This may be a result of underestimated uncertainties on some datasets or rigidities in our model; however, since the overall uncertainties are dominated by the systematic uncertainties described in Section 5, we do not investigate this further here.

\begin{figure*}
 \centering
 \raisebox{0.1cm}{%
\includegraphics[trim=20 20 20 20,clip,width=0.33\textwidth]{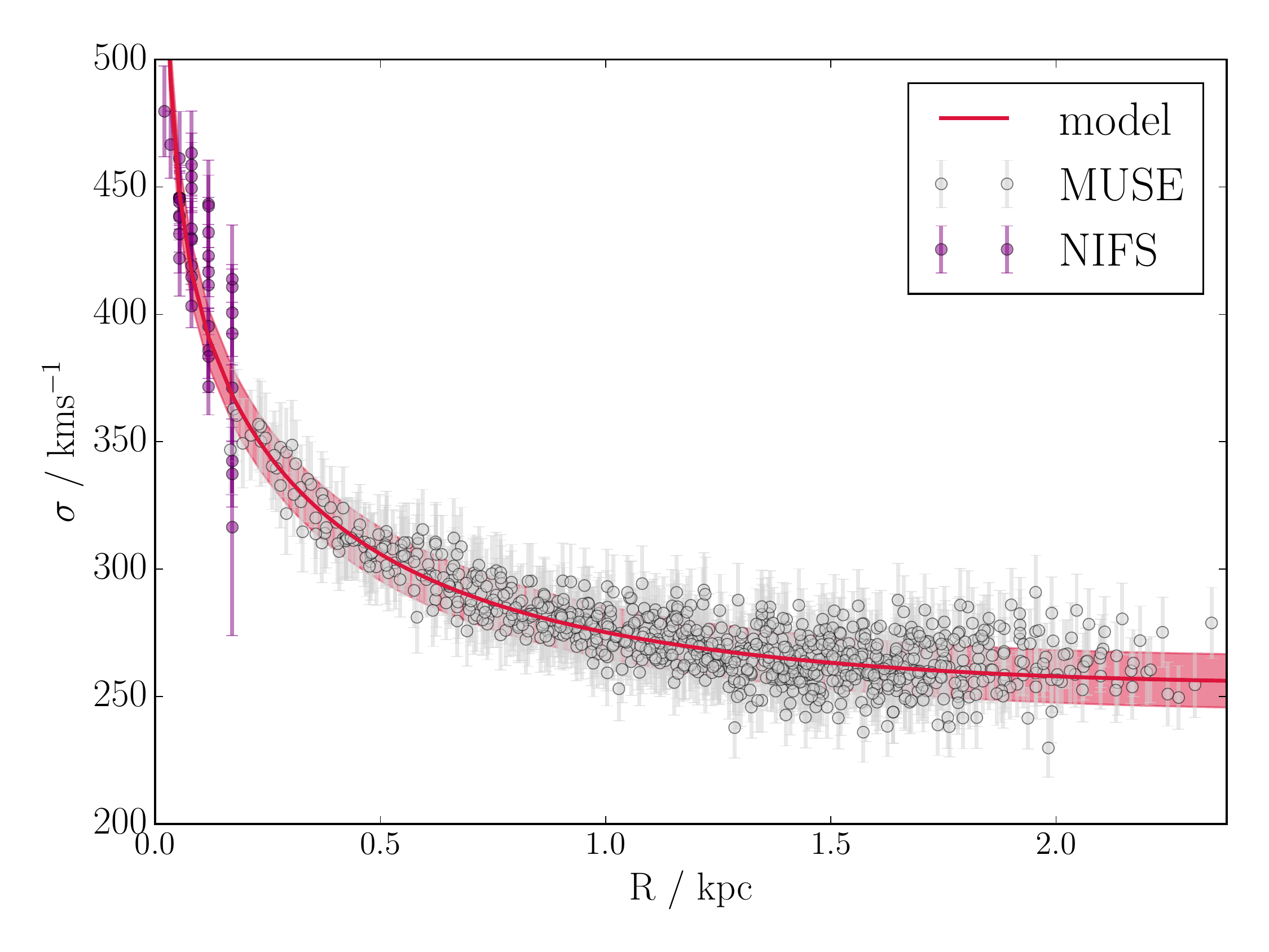}}
\includegraphics[trim=20 20 20 20,clip,width=0.66\textwidth]{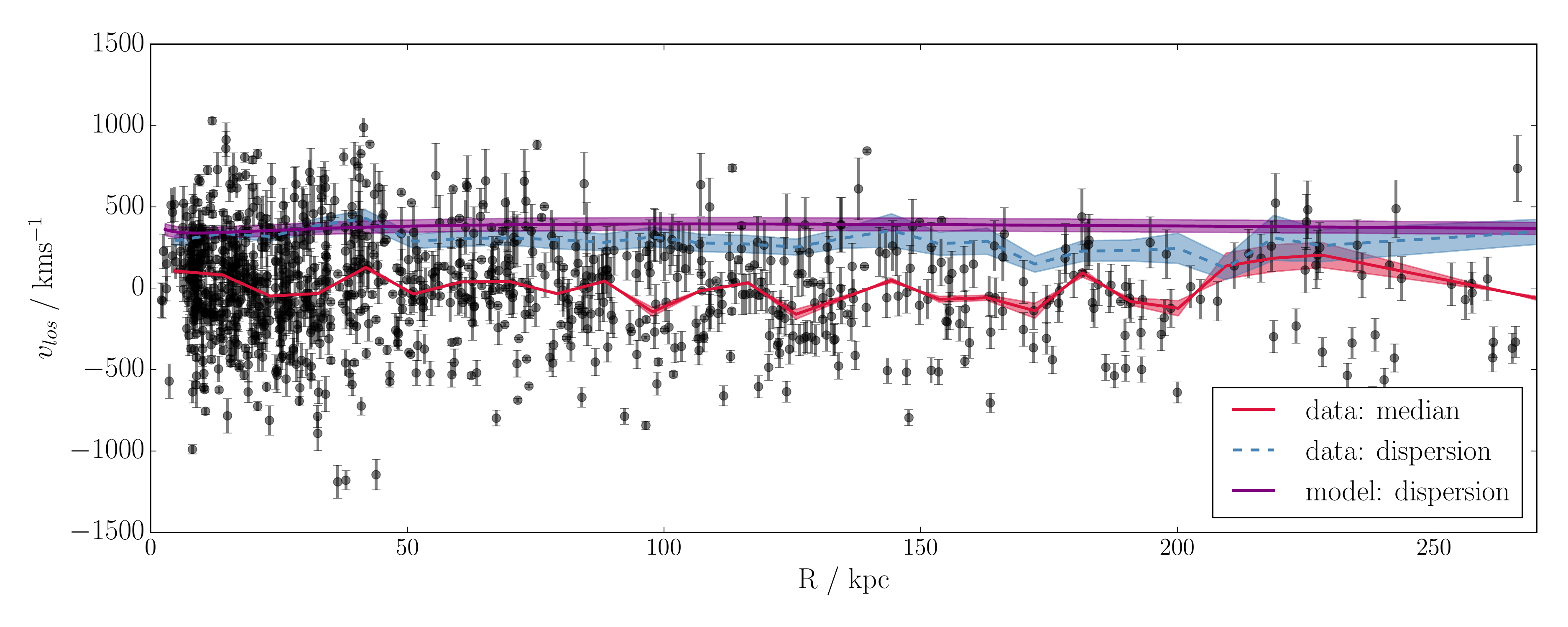}
\caption{Left: Model velocity dispersion profile (red) for the stars as compared with the MUSE (grey) and NIFS (purple) data. Note we are only showing every fifth MUSE datapoint for clarity. Right: Velocity-position plot for the globular clusters, showing datapoints (black), the median and dispersion of the total population as a function of radius (red and blue respectively) and the dispersion profile predicted by our model (purple). Note that our framework does not fit the data dispersion (blue line) directly, and treats the red and blue globular cluster populations separately.}
 \label{fig:chap7fig3new}
\end{figure*}

\begin{figure*}
 \centering
\includegraphics[trim=20 20 20 20,clip,width=0.49\textwidth]{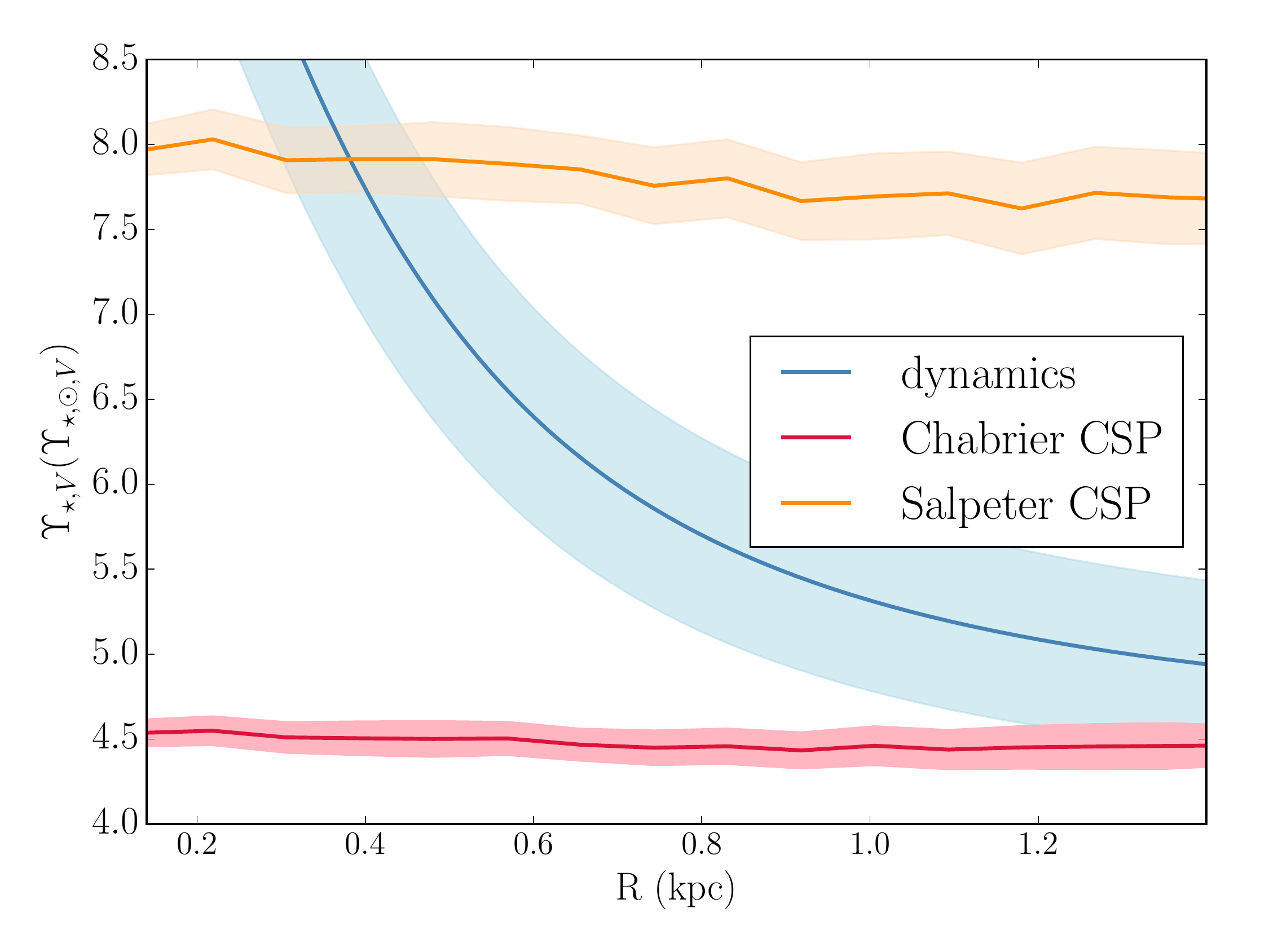} 
\includegraphics[trim=20 20 20 20,clip,width=0.49\textwidth]{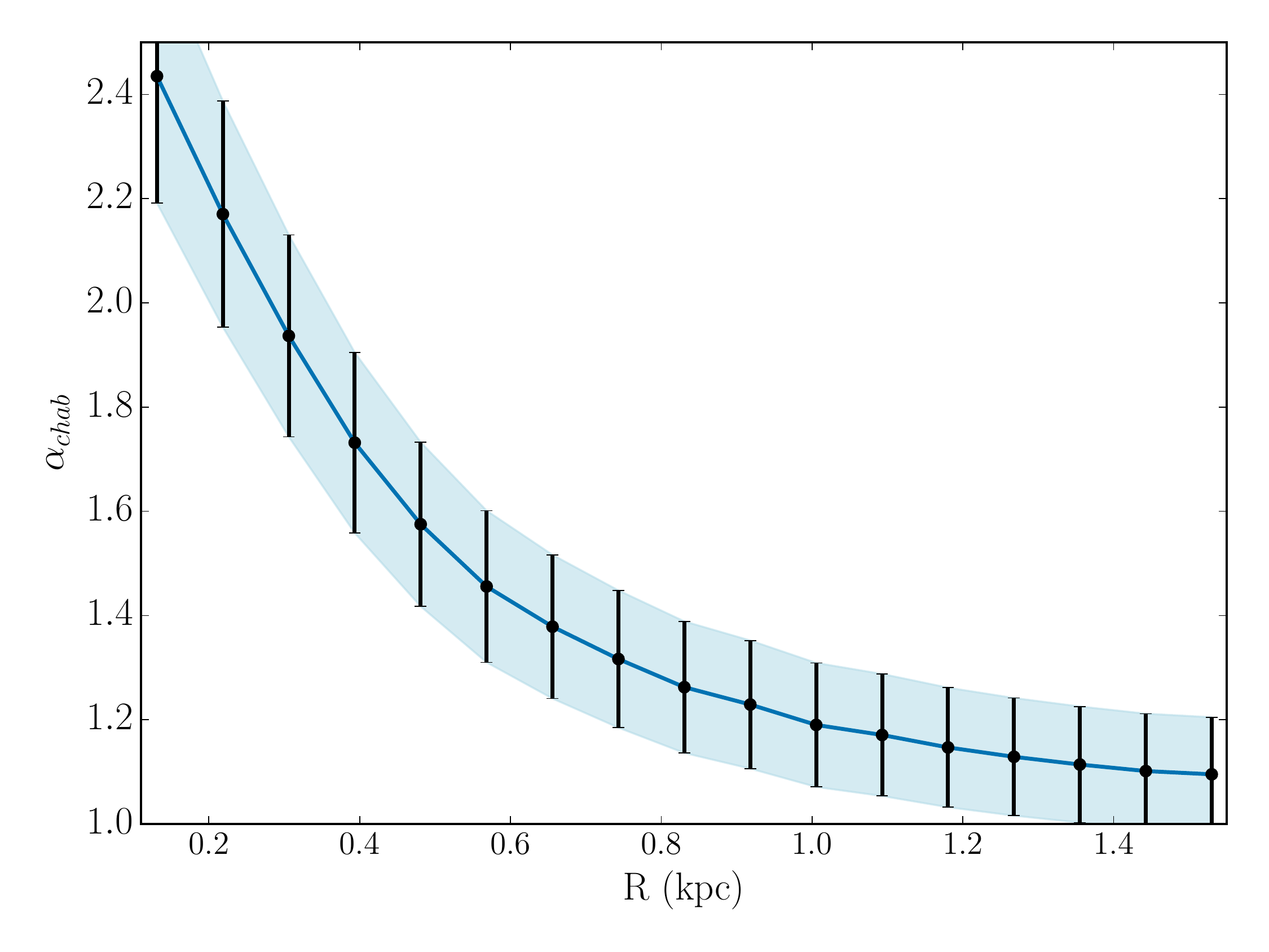}
\caption[Stellar-mass-to-light ratio profile of M87]{Left: The stellar mass-to-light ratio inferred dynamically, shown in blue, declines much more rapidly than can be achieved by gradients in the age, metallicity, star formation history and dust extinction of the stellar populations under a fixed IMF, suggesting that an IMF gradient may be the driving factor. Indeed, the stellar mass-to-light ratio is consistent with stellar population models that assume a Salpeter IMF at small radii, but consistent with stellar population models assuming a Chabrier-like IMF at larger radii. Right: the IMF mismatch-parameter $\alpha_{chab}$ as defined in Equation 15, again showing that the mismatch between the dynamically-inferred stellar mass-to-light ratio and the stellar population-modelling-inferred stellar mass-to-light ratio under the assumption of a Chabrier IMF increases towards the centre of the galaxy. Note that we impose minimum uncertainties of 10\% on $\Upsilon_{\star,V}$ at all radii.}
 \label{fig:chap7fig3}
\end{figure*}

\begin{table*}
 \centering
\begin{tabular}{ccccccccccc}\hline
\\[-1ex]
\multicolumn{11}{c}{MODEL 1: $\displaystyle\Upsilon_{\star}(R) = \alpha \Upsilon_{\star,s} + \frac{\Upsilon_{\star,ch} - \alpha \Upsilon_{\star,s}}{R^2 + R_M^2}R^2$ }\\
\\[-2ex]\hline
model & $\alpha$ & $R_M$ & $\log \rho_0$ & $r_s$ & $\gamma$ & $\beta_{\star}$ & $r_a$ & $\beta_r$ & $\beta_b$ & $M_{BH} \times 10^{-9}$ \\\hline
  I & $>1.48$ & $0.30_{-0.03}^{+0.03}$ & $8.92_{-0.11}^{+0.06}$ & $31.6_{-3.1}^{+4.5}$ & $< 0.25$ & -- & -- & -- & -- & $9.73 \pm 0.97$ \\ 
 A & $>1.48$ & $0.35_{-0.04}^{+0.04}$ & $8.47_{-0.10}^{+0.09}$ & $60.6_{-8.3}^{+10.1}$ & $<0.13$ & $0.92_{-0.09}^{+0.09}$ & $27.6_{-3.3}^{+2.8}$ & $0.34_{-0.23}^{+0.18}$ & $0.32_{-0.16}^{+0.15}$ & $8.04 \pm 0.80$ \\\hline
  \\[-1ex]
 \multicolumn{11}{c}{MODEL 2: $\displaystyle\Upsilon_{\star}(R) = \Upsilon_{\star,1} R^{-\mu}$ }\\
 \\[-2ex]\hline
model & $\Upsilon_{\star,1}$ & $\mu$ & $\log \rho_0$ & $r_s$ & $\gamma$ & $\beta_{\star}$ & $r_a$ & $\beta_r$ & $\beta_b$ \\\hline
I & $4.45_{-0.45}^{+0.45}$ & $-0.55_{-0.05}^{+0.05}$ & $9.47_{-0.05}^{+0.03}$ & $16.87 \pm 1.69$ & $<0.14$ & -- & -- & -- & -- \\
A & $4.34_{-0.45}^{+0.45}$ & $-0.54_{-0.05}^{+0.05}$ & $8.76_{-0.17}^{+0.12}$ & $40.70_{-6.27}^{+9.56}$ & $0.07_{-0.05}^{+0.12}$ & $0.93_{-0.09}^{+0.09}$ & $15.47_{-2.06}^{+2.48}$ & $0.30_{-0.24}^{+0.20}$ & $0.22_{-0.19}^{+0.16}$ \\\hline

\end{tabular}
\caption[Inference on M87's mass allowing a varying stellar-mass-to-light ratio]{Final inference on M87's mass profile for both the isotropic and anisotropic models, with a default M/L profile. Models `I' and `A' are isotropic and anisotropic runs respectively. We report the maximum-posterior values of our samples, along with the 16th and 84th percentiles as a measure of our uncertainty. Where $\gamma$ hits the lower bound of the prior, we give the $95^{th}$ confidence level. All quantities are measured in units of solar mass, solar luminosity, kilometres per second and kiloparsecs. Note that we impose minimum uncertainties of 10\% on all inferred parameters.}
\end{table*}

\section{Stellar population modelling}
\label{sec:chap7sec6}

The key result of the dynamical modelling presented in this study is that the stellar mass-to-light ratio of M87 is a declining function of radius. The stellar-mass-to-light ratio as measured dynamically represents the summed contributions to both the mass and the light from across the stellar population(s) and so is sensitive to the age and metallicity of those populations, in addition to the integral over their mass function.  From the dynamical inference alone, it is not possible to identify the driving factor behind the gradient that we infer.

To disentangle the contributions to the mass-to-light ratio due to these different stellar properties, we carry out stellar population modelling using the high-resolution, extinction-corrected eleven-band HST photometry of the central $17.5''$ $(=1.4$ kpc) of M87 that was presented by \citet{Montes2014}. These data span a wide range of filters from F336W (HST/WFPC2) to F850LP (HST/ACS), and are presented in that paper as surface brightness measurements within circular annuli of width $1''$. We can therefore use stellar population models to infer the age, metallicity, star formation history, dust extinction and stellar mass-to-light ratio as a function of projected radius, making this the ideal dataset to compare with our dynamical inference.

We model the photometry following the methods of \citet{Auger2009,Oldham2017}. We use the composite BC03 stellar population models \citep{Bruzual2003} to compute apparent magnitudes in the 11 filters on a grid of stellar age $T$, metallicity $Z$, dust extinction $\tau_V$ and time constant $\tau$ of an exponentially decaying star formation history, and construct a spline interpolation model which allows magnitudes to be evaluated at any point within the grid; these magnitudes can then be scaled by the stellar mass. We then explore the posterior probability distribution of these parameters, along with the stellar mass, using \textsc{Emcee}. We treat each radius `bin' as completely independent, such that $Z$, $T$, $\tau$, $\tau_V$ and $M_{\star}$ can vary freely as a function of radius, and impose a `global' IMF that is either Chabrier or Salpeter.

The resulting stellar-mass-to-light ratio profiles that we infer therefore tell us not only the difference in magnitude of the stellar mass-to-light ratio under different IMF assumptions, but also the stellar mass-to-light ratio slope that can be achieved by allowing gradients in all parameters \emph{except} for the IMF. Figure A5 in the Appendix shows an example of our inference on the various stellar population properties for a Chabrier IMF: the gradients in metallicity and age are small, resulting in a significantly shallower stellar mass-to-light ratio gradient than that inferred dynamically (Figure~\ref{fig:chap7fig3}); the result for a Salpeter IMF is qualitatively the same. It appears, then, that age, metallicity and star formation history variations cannot be driving the $\Upsilon_{\star}$ slope, and radial gradients in some other property must be responsible for this. A radially varying IMF, falling from a Salpeter-like function to a more Chabrier-like one with increasing radius, may instead be the driving factor.

\begin{figure}
 \centering
\includegraphics[trim=20 20 20 20,clip,width=0.49\textwidth]{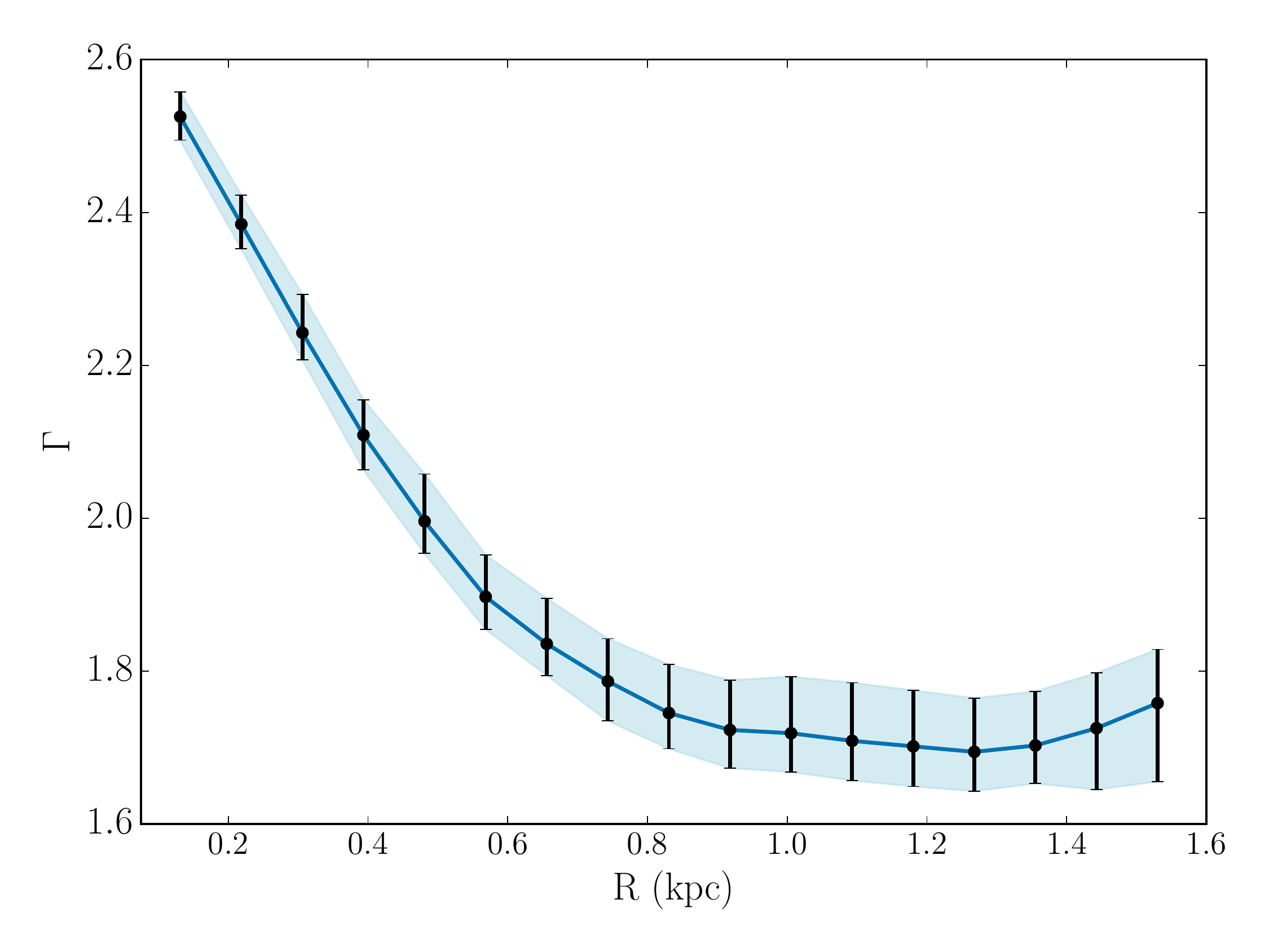}
\caption[Inference on M87's IMF slope]{Inference on the slope of a broken-power-law IMF with slope $\xi = 2.3$ for $M>1 M_{\odot}$ and $\xi = \Gamma$ for $M<1M_{\odot}$. We use FSPS to calculate magnitudes on a grid of ages, metallicities and $\Gamma$, and obtain the posteriors on these quantities based on 11-band HST photometry and our dynamical inference on the projected stellar mass as a function of radius. Our model clearly requires an IMF that becomes increasingly bottom-heavy towards the centre. Nevertheless, it is possible that alternative models which allow greater flexibility in other stellar population properties may also be able to reproduce the photometry and mass inference simultaneously. }
 \label{fig:chap7fig4}
\end{figure}

Based on these findings, we attempt to \emph{infer} the IMF slope as a function of radius by constructing stellar population models which are required to fit the photometry and our inference on the projected mass profile simultaneously. We use the Flexible Stellar Population Synthesis (FSPS) models of \citet{Conroy1,Conroy2}, which allow significantly more freedom in the form of the IMF than the BC03 machinery. We compute magnitudes on a grid of metallicities, ages and low-mass IMF slopes $\Gamma$, where for the IMF we assume a double power-law with the form
\begin{equation}
\frac{d\log N}{d\log m} \propto m^{-\xi}, \begin{cases} \xi = \Gamma & \mbox{$m<1M_{\odot}$} \\
 \xi = 2.3 &\mbox{$m>1M_{\odot}$} \end{cases}
\end{equation}
such that the IMF follows the canonical (e.g. Chabrier, Kroupa, Salpeter) form at high masses, but is flexible at the low-mass end to allow the data to choose between bottom-heaviness ($\Gamma > 2.3$) and bottom-lightness ($\Gamma < 2.3$). Note that $\Gamma = 2.3$ corresponds to a Salpeter IMF and $\Gamma = 1.3$ corresponds to an IMF which is Milky-Way-like. Guided by the BC03 analysis, in which $\tau_V$ and $\tau$ change negligibly with radius (and $\tau$ is short), we consider single stellar populations with a fixed dust extinction parameterised as in \citet[][similarly to BC03]{Charlot2010}, and allow $Z$, $T$ and $\Gamma$ to vary freely with radius. We then require our model to reproduce the photometry under the condition that the mass profile follows that which we have inferred dynamically. Figure~\ref{fig:chap7fig4} shows our inference on $\Gamma$ as a function of radius: we find that the IMF slope implied by the photometry and dynamics is super-Salpeter in the innermost radial bins and becomes approximately Milky-Way-like by the outermost bin. We find universally old stellar ages $\log T\sim 10.07$ Gyr and consistently supersolar metallicities; both agree with our BC03 anaylsis. This model represents one possible explanation for M87's stellar mass-to-light ratio gradient, but we cannot rule out the possibility that more flexible stellar population modelling, allowing freedom in a greater number of parameters, could also explain our result.

Finally, we note that the MUSE spectra for M87 are suitable for more thorough stellar population modelling based on spectral features and should allow direct inference on the IMF as a function of radius; a comparison of our dynamical inference with such an analysis will be presented in the forthcoming work of \citet{Sarzi2017}. 

\begin{figure*}
 \centering
\includegraphics[trim=20 20 20 20,clip,width=0.49\textwidth]{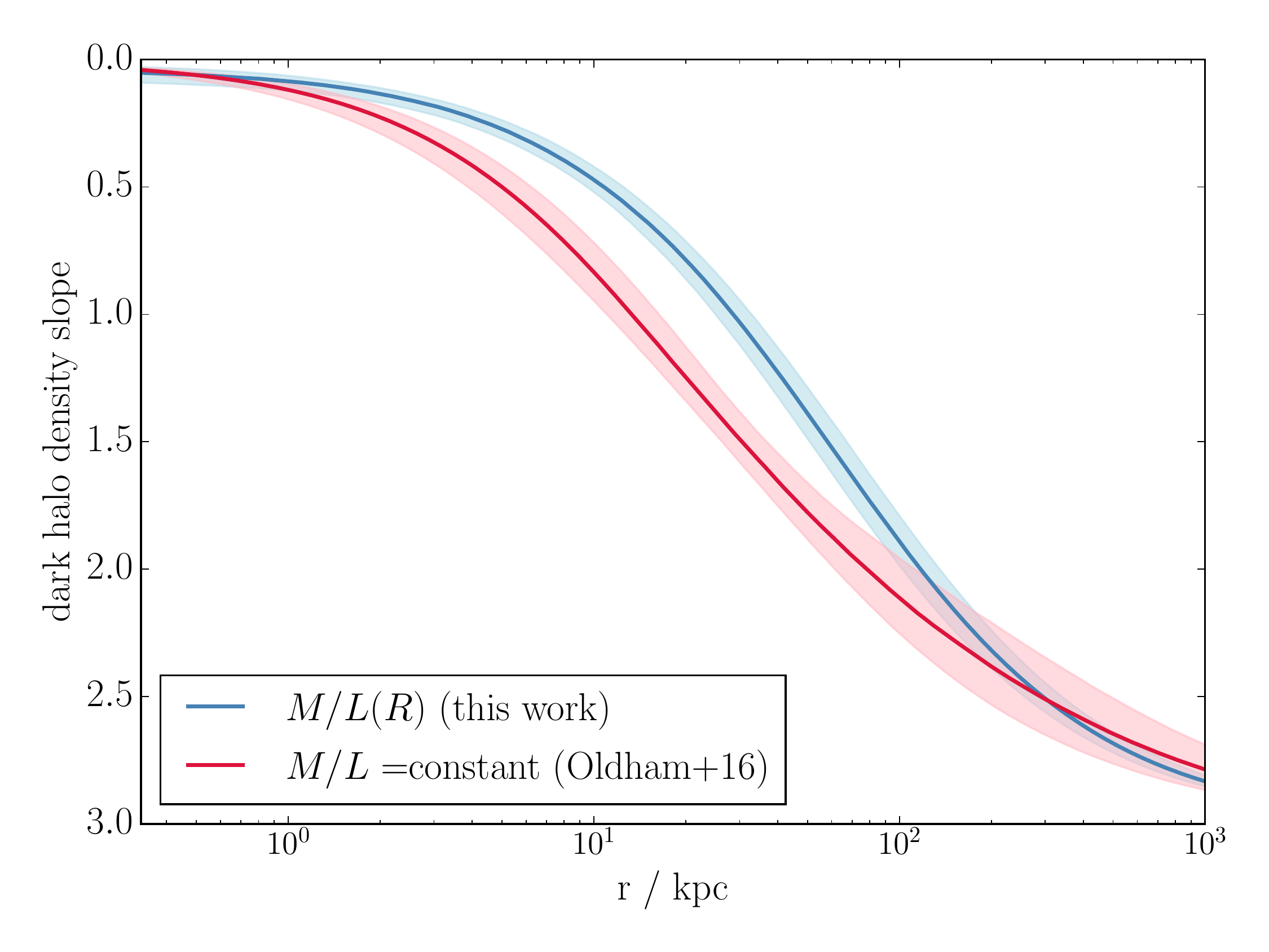} 
\includegraphics[trim=20 20 20 20,clip,width=0.49\textwidth]{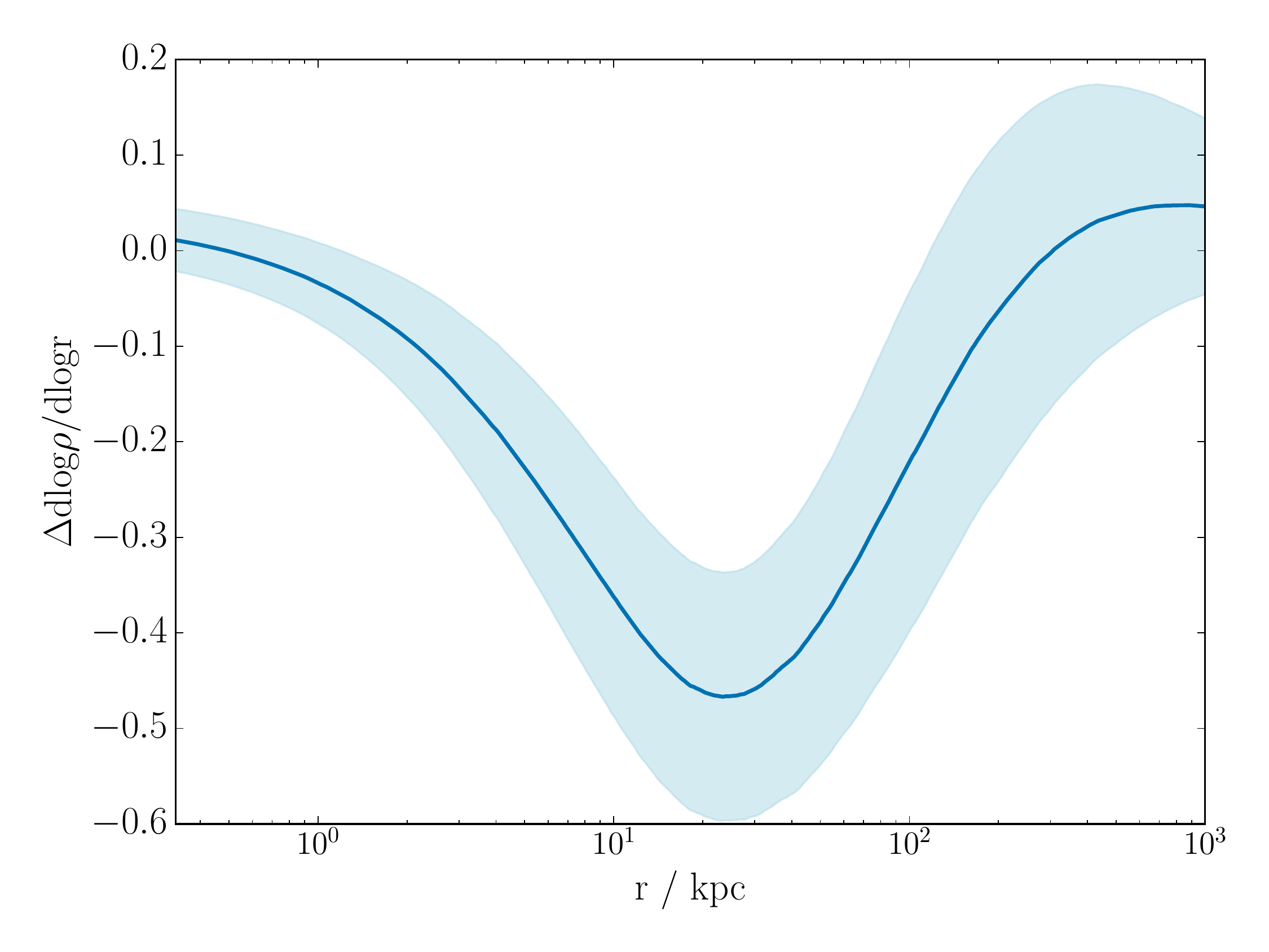} 
\caption[Inference on M87's logarithmic halo density slope]{Left: Inference on the dark halo density slope (defined as $\gamma$ in Equation 2) as a function of 3D radius in this work (blue) and the work of \citet{Oldham2016b} (red), in which a constant stellar mass-to-light ratio was assumed. Right: Difference in slope between this work and \citet{Oldham2016b} (i.e. blue curve minus red curve from the left-hand panel) as a function of radius. The increased slope that is allowed for the stellar mass in this work leads to a larger core in the halo structure at small radii, though leaving the qualitative conclusions unchanged.}
 \label{fig:chap7fig5}
\end{figure*}

\section{Discussion}
\label{sec:chap7sec7}

By dynamically modelling the stars, globular clusters and satellite galaxies in M87's gravitational potential, we have disentangled the contributions from the dark and stellar mass to show that (a) the dark halo is centrally cored, and (b) the stellar IMF is a declining function of radius. In the following sections, we discuss these two results in more detail.

\subsection{Inference on the halo structure is robust}
\label{sec:chap7sec7sub1}
This study serves as an extension of the models of \citet{Oldham2016b}, which set the first constraints on M87's dark matter core. There, we investigated a number of different parameterisations for both the halo and the anisotropy profiles, and showed that the inference on the halo structure was robust against reasonable (and computationally feasible) changes in the model. However, we did not investigate the effect of our assumption of a constant stellar mass-to-light ratio, and this remained a dominant source of systematic uncertainty in our result. In this work, we have removed this source of uncertainty by showing that allowing stellar mass-to-light ratio gradients, whilst affecting the stellar mass profile and the finer details of the halo profile, does not remove the need for a central dark matter core. In Figure~\ref{fig:chap7fig5}, we show a comparison between the inferred dark halo density slope from \citet{Oldham2016b} and the present work. The core we find here is larger: as shown in Figure~\ref{fig:chap7fig5} (right), we find a shallower slope $\Delta \gamma \leq 0.6$ at intermediate radii, though some of this is a consequence of the analytic profiles that we impose for both the dark matter and the stellar mass, which limit the freedom that the slope is allowed at different radii. Despite this variation, it is nevertheless clear that the \emph{qualitative} nature of the halo structure is robust against variations in the stellar mass-to-light ratio model; M87 does indeed appear to have less dark matter centrally than the NFW prediction, indicating the action of baryonic physics in the form of AGN feedback or dynamical friction during satellite infall \citep[e.g.][]{Laporte2012,Martizzi2012}. The physical implications of this are discussed more extensively in \citet{Oldham2016b}; here, we simply note that the heating of the halo by satellite accretion would be strikingly consistent with the interpretation of the IMF gradients that we suggest below.

\subsection{Radial gradients in the IMF?}

This study represents the first use of dynamical tracers to disentangle the mass contributions from a dark matter halo and a stellar component with a mass-to-light ratio gradient. In our main analysis, we implement two independent models for the stellar mass-to-light ratio, both of which robustly prefer a stellar mass-to-light ratio which declines strongly within the central $\sim 1$ kpc; furthermore, we find that more general models also reproduce this result. Stellar population modelling of high-resolution multiband photometry suggests that this gradient is much stronger than can be achieved by gradients in either metallicity, age, extinction or star formation history, and therefore that a varying IMF -- transitioning from a heavy Salpeter-like IMF in the very centre to a lighter Milky-Way-like IMF within a stellar effective radius -- may be responsible. Whilst we cannot rule out the possibility that this excess central mass could be due to non-stellar material such as intermediate-mass black holes or neutron stars, the agreement of our results with stellar population studies in other massive ETGs is compelling. In \citet{MartinNavarro2015}, stellar population modelling was used to measure radial gradients in the IMF slope in two high-mass ETGs -- treating the IMF as a single power law which tapers off to a constant value below $M<0.6M_{\odot}$ -- and found that the slope at the effective radius was around half its value in the centre, implying a significant excess of low-mass stars in the central regions. Since then, \citet{vanDokkum2017} have found evidence for qualitatively similar trends in six other massive ETGs, which appear to be generally well-described as having centrally bottom-heavy IMFs which become Milky-Way-like beyond the central $\sim 0.4R_e$. Moreover, \citet{Sarzi2017} performed stellar population modelling of the same MUSE spectra used in this paper to derive a consistent result with our own. Whilst the dynamical modelling presented in this study is sensitive to a different aspect of the IMF -- that is, the integral over the IMF, rather than the fraction of low-mass stars -- the radially-declining stellar mass-to-light ratio that we find can be naturally explained in terms of a decreasing fraction of low-mass stars. These independent and consistent results from dynamics and stellar population modelling therefore make a compelling case for the existence of a fundamentally different, bottom-heavy IMF in the innermost regions of massive ETGs. 

If correct, this dependence of the form of the IMF on galaxy radius has important implications for our understanding of ETG assembly and evolution. First, that all but the very innermost regions are consistent with having Milky-Way-like IMFs suggests that the star formation conditions in these regions do not differ significantly from those found in lower-mass galaxies. In the context of the recent accumulating evidence for the build-up of massive ETGs being dominated by repeated minor mergers and accretion \citep[e.g.][]{vanDokkum2010b}, this may be the result of a large amount of the larger-radius stellar material being originally formed in lower-mass Milky-Way-like systems which were subsequently accreted. On the other hand, the bottom-heavy IMFs in the cores of these ETGs imply fundamentally different star formation conditions in these regions; as these galaxies most likely form in the centres of the most massive dark haloes with the deepest potential wells, these different conditions may arise from the especially dense environments in which initial star formation must take place. In this picture -- in which the compact central core forms at early times with a bottom-heavy IMF, followed by the ongoing accretion of lower-mass satellites with less bottom-heavy IMFs -- the gradient that we infer here arises naturally, and may also be expected to be particularly pronounced in M87, which sits at the high-mass end of the galaxy mass function, in the centre of a cluster, and surrounded by a vast population of globular clusters. (The same explanation may also account for our inference on the dark matter structure, via the heating of the halo by dynamical friction from infalling satellites; \citealp{Laporte2012}.) We note that the result of \citet{Dutton2013} that bulge-dominated late-type galaxies require Salpeter-like IMFs in their central bulge regions, but not necessarily at larger radii, may be related to similar growth mechanisms.  An important future step in understanding the origin of these IMF gradients will be to examine their strength across the ETG mass function.

\subsection{The importance of accounting for IMF gradients}

M87 is a massive, nearby BCG for which extensive data are available, such that it is possible for us to construct models which constrain the gradient in its stellar mass-to-light ratio. However, for the majority of ETGs for which stellar mass-to-light ratios are measured, limitations in the data may make obtaining such constraints difficult. It is therefore important to consider the systematic uncertainties that are introduced by modelling a varying stellar mass-to-light ratio as constant. Given the previous work on M87 presented in \citet{Oldham2016b}, we are now in a position to do this.

First, we note from Section~\ref{sec:chap7sec7sub1} that variations in the stellar mass model do not affect our inference on the halo structure on a qualitative level (though the strength of the core does change; see Figure 5). This is encouraging as it further justifies previous attempts to disentangle the dark and luminous mass in ETGs in order to infer their halo structure \citep[e.g.][]{Newman2015}. Nevertheless, M87 has a very low central dark matter fraction due to the cored nature of its halo, and it is possible that this may minimise the degeneracy between the dark and luminous mass. ETGs in more isolated environments than M87 may have significantly cuspier haloes (\citealp{Sonnenfeld2012}); it is possible that the existence of stellar mass-to-light ratio gradients in these systems may have a greater impact on the inference on the halo structure. This is an issue that will need to be addressed in the future; the recently-discovered low-redshift lenses \citep{Smith2015}, which reside in a range of environments and for which both lensing and extensive dynamical information is accessible, may represent suitable opportunities for this.

In terms of the stellar mass, however, we note that our difficulty in breaking the mass-anisotropy degeneracy in the stellar mass in our radially-constant-$\Upsilon_{\star}$ models -- with the stellar mass changing significantly depending on the complexity of our anisotropy model -- may be a result of the inadequacy of the constant-$\Upsilon_{\star}$ model. Indeed, when we implement this more flexible radially-varying $\Upsilon_{\star}$ model, our inference on the mass structure of both the dark and stellar mass agree regardless of our assumptions about the anisotropy. The implication is that care must be taken in interpreting the stellar mass that is inferred when a constant stellar mass-to-light ratio is assumed, and attention paid to which particular aspects of the data are driving the inference. Furthermore, we note that the isotropic model of \citet{Oldham2016b} required a significantly larger stellar mass-to-light ratio. The role played by this source of systematic uncertainty in the correlations of the IMF mismatch parameter with other galaxy properties \citep[such as the stellar velocity dispersion; see e.g.][]{Auger2010b} also needs to be better understood. Finally, we emphasise that the existence of IMF gradients complicates the comparison of measurements in different galaxies based on data extracted over different physical apertures, and that care must be taken to ensure that meaningful comparisons are made.

\section{Summary and conclusions}
\label{sec:chap7sec8}

We have combined multiple kinematic tracers of the mass in M87 to disentangle the dark and luminous mass components and the stellar anisotropy, and have inferred the presence of stellar mass-to-light ratio gradients in this massive galaxy for the first time. Our main conclusions are summarised below.
\begin{enumerate}
\item The stellar mass-to-light ratio $\Upsilon_{\star,v}$ in M87 is a declining function of radius. Parameterising $\Upsilon_{\star,v}$ in the $V$-band by a central mismatch parameter relative to a Salpeter IMF, $\alpha$, and a scale radius $R_M$ governing the transition to a Chabrier-like IMF, we find $\alpha>1.48$ at 95$\%$ confidence and $R_M = 0.35 \pm 0.04$ kpc. Modelling $\Upsilon_{\star,v}$ as a power law, we find a slope $\mu = -0.54 \pm 0.05$ and $\Upsilon_{\star,v} = 4.34 \pm 0.43$ at 1 kpc from the centre.
\item Multi-band, high-resolution photometry indicates that such a strong stellar mass-to-light ratio gradient cannot be achieved by varying only the metallicity, age, dust extinction and star formation history of the stellar population if the IMF remains fixed. On the other hand, the stellar mass-to-light ratio gradient that we infer is consistent with M87 having an IMF which is Salpeter-like in the central $\sim 0.5$ kpc and becomes Chabrier-like at $\sim 3$ kpc.
\item Inference on the halo structure does not change qualitatively depending on whether or not these stellar mass-to-light ratio gradients are allowed for in the model, though the size of the core changes significantly. Moving forward, it will be important to account for the presence of stellar mass-to-light ratio gradients when interpreting stellar mass measurements and their implications for the non-universality of the IMF in ETGs.
\item Taken together, M87's dark matter core and radially-varying IMF can be explained coherently if the galaxy formed its central stellar populations in un-Milky-Way-like physical conditions -- leading to the stellar mass in this region having a bottom-heavy IMF -- and subsequently accreted material from more Milky-Way-like systems at larger radii, in a process which both ingrained a stellar-mass-to-light ratio gradient and dynamically heated the halo to create a dark matter core.
\end{enumerate}

\section{Acknowledgements}

We thank the referee for helpful feedback which improved the paper. LJO thanks the other participants in the IMF workshop in Leiden for constructive discussions, Marc Sarzi for sharing his stellar population modelling results, and Cameron Lemon for useful comments on the manuscript. Both authors thank the Science and Technology Facilities Council (STFC) for financial support in the form of a studentship (LJO) and an Ernest Rutherford Fellowship (MWA).

\newpage
\onecolumn
\appendix
\section{Inference on M87's mass structure}

\noindent\begin{minipage}{\textwidth}
 \centering
\includegraphics[trim=70 70 70 70,clip,width=\textwidth]{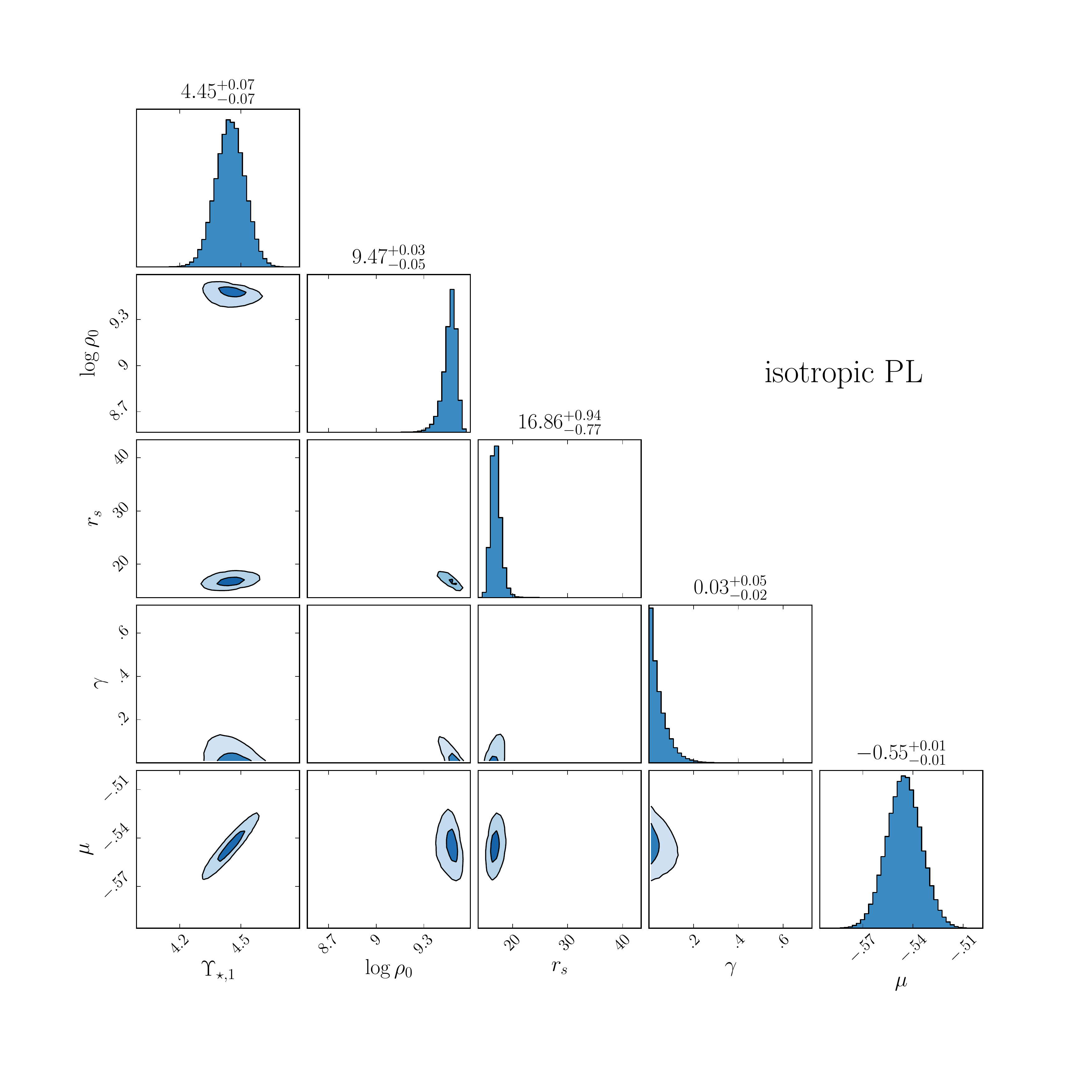}
\captionof{figure}{Inference on the dynamical model under assumptions of isotropy for all tracer populations and the PL $\Upsilon_{\star}$ model. Units are the same as in Table 1.}
\end{minipage}

\begin{figure*}
 \centering
\includegraphics[trim=70 70 70 70,clip,width=\textwidth]{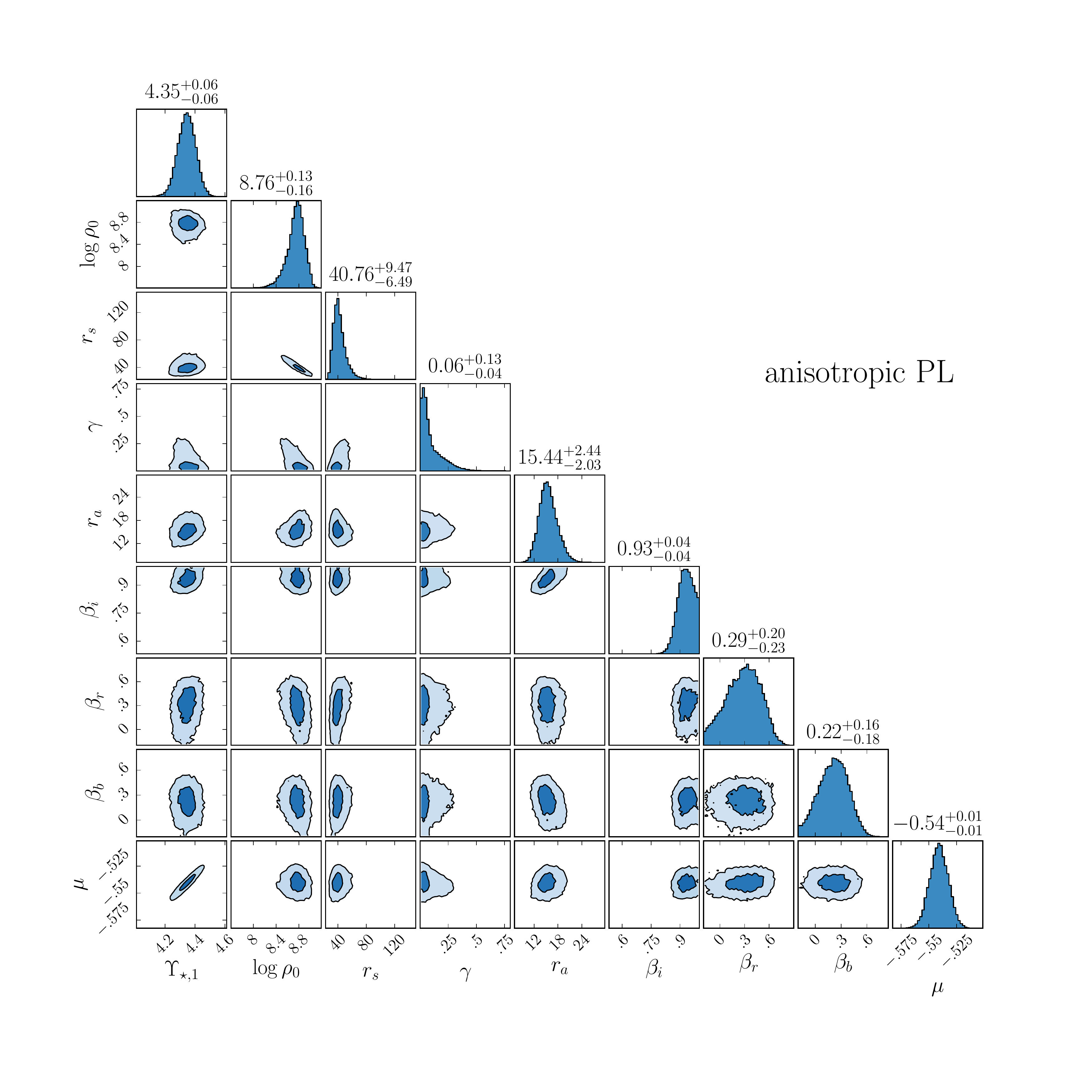}
\caption[Inference on anisotropic PL model of M87's mass]{Inference on the dynamical model under assumptions of anisotropy for all tracer populations and the PL $\Upsilon_{\star}$ model, as detailed in Section 3. Units are the same as in Table 1.}
\end{figure*}

\begin{figure*}
 \centering
\includegraphics[trim=70 70 70 70,clip,width=\textwidth]{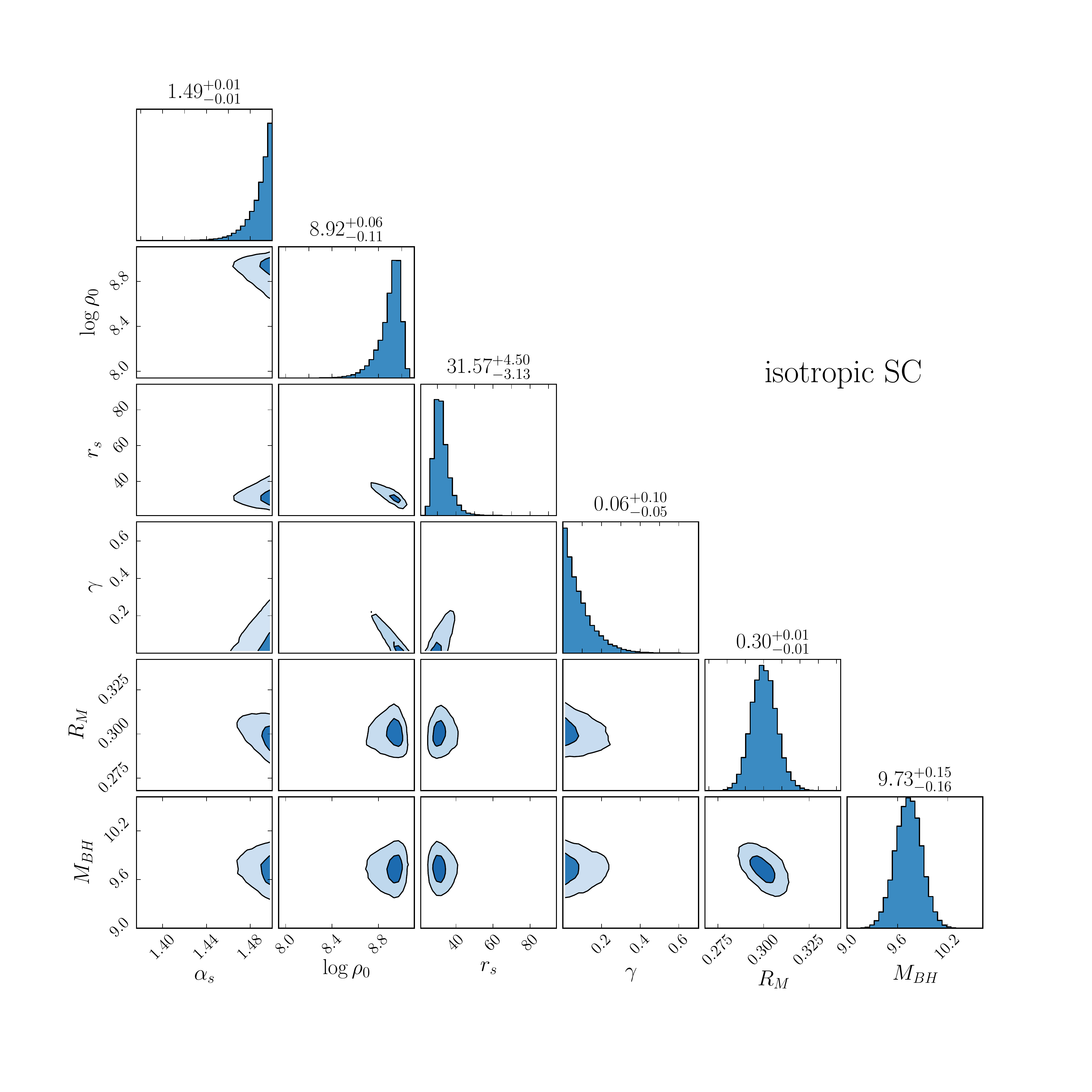}
\caption[Inference on isotropic SC model of M87's mass]{Inference on the dynamical model under assumptions of isotropy for all tracer populations and the SC $\Upsilon_{\star}$ model. Units are the same as in Table 1.}
\end{figure*}

\begin{figure*}
 \centering
\includegraphics[trim=70 70 70 70,clip,width=\textwidth]{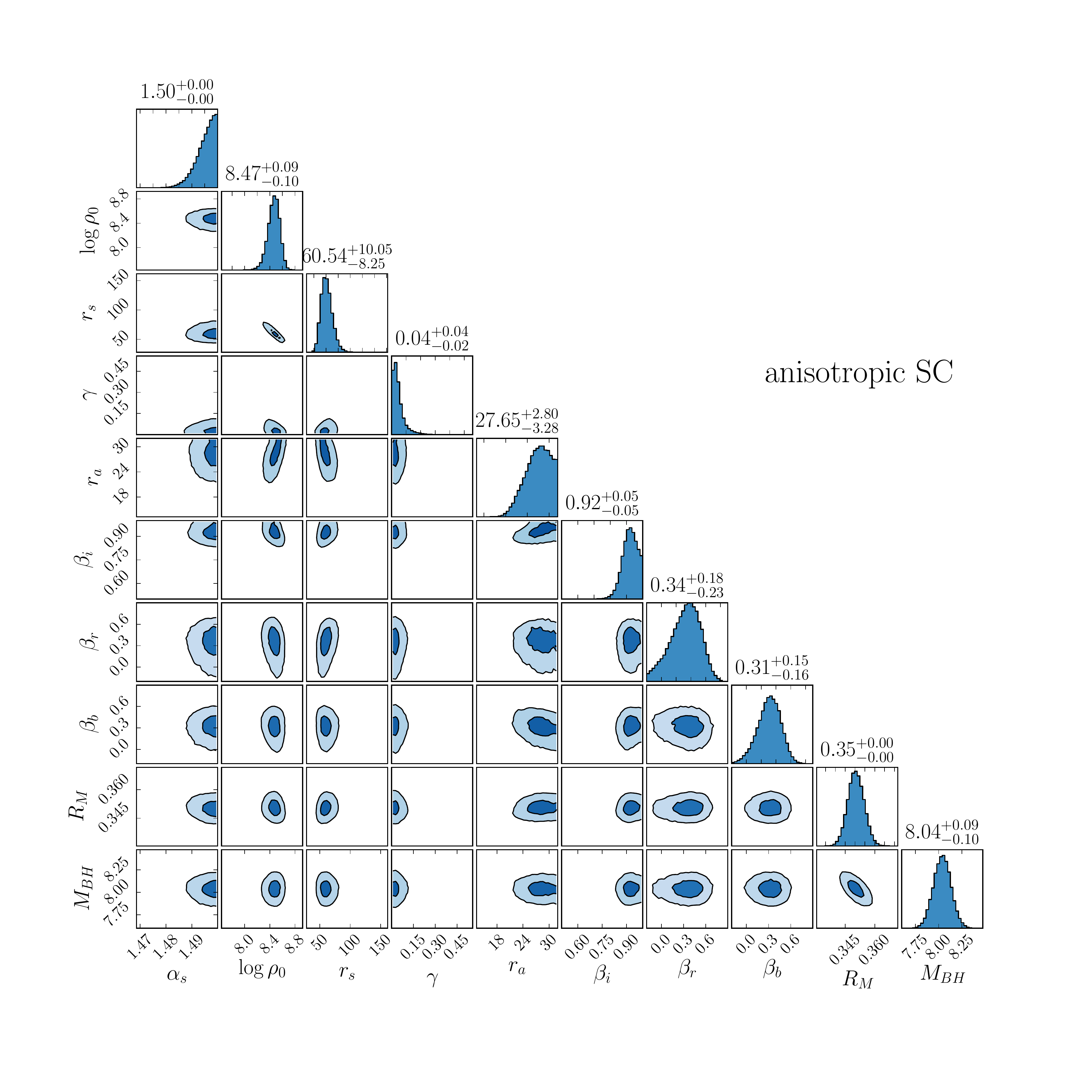}
\caption[Inference on anisotropic SC model of M87's mass]{Inference on the dynamical model under assumptions of anisotropy for all tracer populations and the SC $\Upsilon_{\star}$ model, as detailed in Section 3. Units are the same as in Table 1.}
\end{figure*}

\begin{figure*}
 \centering
\includegraphics[trim=20 20 20 20,clip,width=\textwidth]{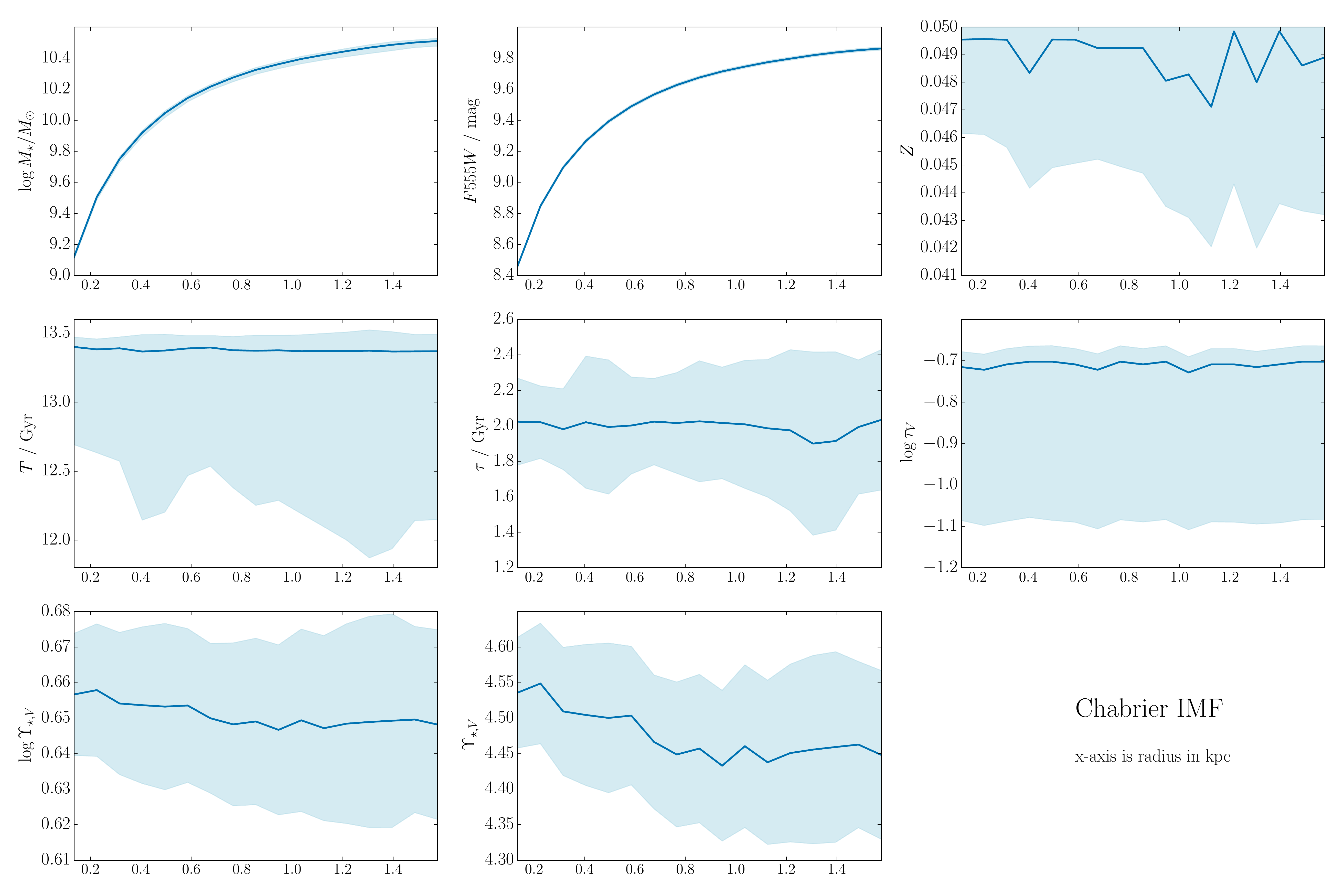}
\caption[Inference on M87's stellar population properties]{Inference on stellar population properties, modelling high-resolution 11-band HST photometry using the models of BC03. Allowing for gradients in all parameters except the IMF allows only very weak gradients in the stellar mass-to-light ratio, suggesting that the main cause of the stellar mass-to-light ratio gradient that we infer dynamically may be IMF variations. This Figure shows our inference assuming a Chabrier IMF; our conclusions are qualitatively the same when a Salpeter IMF is assumed instead.}
\end{figure*}

\end{document}